\documentclass[aps,prb,twocolumn,10pt,superscriptaddress]{revtex4-2} 
\usepackage[utf8]{inputenc}
\usepackage[T1]{fontenc}
\usepackage{amsmath}
\usepackage{enumerate}
\usepackage{xcolor,colortbl}
\usepackage{soul}
\usepackage{multirow}
\usepackage{graphicx}
\usepackage{dsfont}
\usepackage{physics}
\usepackage{hyperref}
\usepackage{stmaryrd}
\hypersetup{
    colorlinks=true,
    linkcolor=purple,
    citecolor=teal,     
    urlcolor=blue,
    pdftitle={},
}
\usepackage{orcidlink}
\usepackage{comment}
\usepackage{adjustbox}
\newcommand{\unit}[1]{\ensuremath{\,\mathrm{#1}}}

\begin{document}

\title{Compiling the surface code to crossbar spin qubit architectures}

\author{D\'{a}vid Pataki\,\orcidlink{0000-0002-8818-3723}}
%\email{email: pataki.david@ttk.bme.hu}
\affiliation{Department of Theoretical Physics, Institute of Physics, Budapest University of Technology and Economics, Műegyetem rkp. 3., H-1111 Budapest, Hungary}

\author{Andr\'{a}s P\'{a}lyi}
\affiliation{Department of Theoretical Physics, Institute of Physics, Budapest University of Technology and Economics, Műegyetem rkp. 3., H-1111 Budapest, Hungary}
\affiliation{HUN-REN--BME Quantum Dynamics and Correlations Research Group, Műegyetem rkp. 3., H-1111 Budapest, Hungary}

\begin{abstract}
Spin qubits in quantum dots provide a promising platform for realizing large-scale quantum processors since they have a small characteristic size of a few tens of nanometers. One difficulty of controlling e.g., a few thousand qubits on a single chip is the large number of control lines. The crossbar control architecture has been proposed to reduce the number of control lines exploiting shared control among the qubits. Here, we compile the surface code cycle to a pulse sequence that can be executed in the crossbar architecture.
We decompose the quantum circuits of the stabilizer measurements in terms of native gates of the spin-qubit architecture. We provide a routing and scheduling protocol, and construct a gate voltage pulse sequence for the stabilizer measurement cycle. During this protocol, coherent phase errors can occur on idle qubits, due to the operational constraints of the crossbar architecture. 
We characterize these crosstalk errors during the stabilizer measurement cycle, and identify an experimentally relevant parameter regime where the crosstalk errors are below the surface code threshold. 
Our results provide design guidelines for near-term qubit experiments with crossbar architectures.
\end{abstract}

\maketitle

\tableofcontents

\section{Introduction}

Quantum computers have the potential to solve problems that are intractable for classical computers, but they are highly susceptible to errors due to decoherence and other noise sources. This makes error correction crucial to achieve practical quantum computation.
One of the most prominent quantum error correction codes is the surface code \cite{Fowler_2012}, 
which is based on a two-dimensional qubit grid and requires only nearest-neighbor connectivity.

The surface code is a code family, whose members are labelled by the odd integer $d$, called the code distance. 
The distance-$d$ surface code encodes a single logical qubit, and it has code size $d^2$, meaning that it consists of $d^2$ physical data qubits. 
The surface code shows \emph{threshold behavior}, which means that if the strength of physical errors is below a threshold, then the code performs increasingly better as the code size is increased, with vanishing logical error for asymptotically large code size \cite{Dennis_2002,Fowler_2012}. The error threshold of the surface code is around $1\%$ for realistic (circuit-level) noise which is within reach for state-of-the-art solid-state quantum hardware, e.g. superconducting qubits \cite{Krinner_2022,Zhao_2022,GoogleQuantum_2023,GoogleQuantum_2024}.

In recent years, semiconductor spin qubits have emerged as a promising platform for future quantum computers.
A single electron or a single hole confined in a semiconductor quantum dot has a small characteristic size (10-100 nm) \cite{Hendrickx_2021,Burkard_2023} compared to superconducting qubits (100 microns) \cite{Kjaergaard_2020}. Moreover, spin qubits also have high-fidelity universal quantum control (single-qubit and two-qubit gate fidelities above $99.5\%$) \cite{Noiri_2022_universal,Xue_2022}, the capability of operation above one kelvin \cite{Vandersypen_2017,Petit_2020,Yang_2020,Camenzind_2022,Huang_2024}, and their fabrication exploits today's highly advanced technologies of the semiconductor industry \cite{Maurand,Veldhorst_2017,Neyens}.
Several architectural proposals have been made for scaling up such spin qubit platforms \cite{Zajac_2016,Veldhorst_2017,Nigg_2017,Buonacorsi_2019,Boter_2019,Boter_2022,patomäki2023pipeline,Künne_2024}.
One difficulty of controlling, e.g., a few thousand qubits on a single chip is a large number of control lines: this number scales linearly with the qubit count. A crossbar control architecture was proposed to overcome this issue, where the number of control lines scales as the square root of the qubit count \cite{Li_2018}, making it a good candidate for realizing large-scale quantum computers.

\begin{figure}
    \centering
    \includegraphics[width=0.45\textwidth]{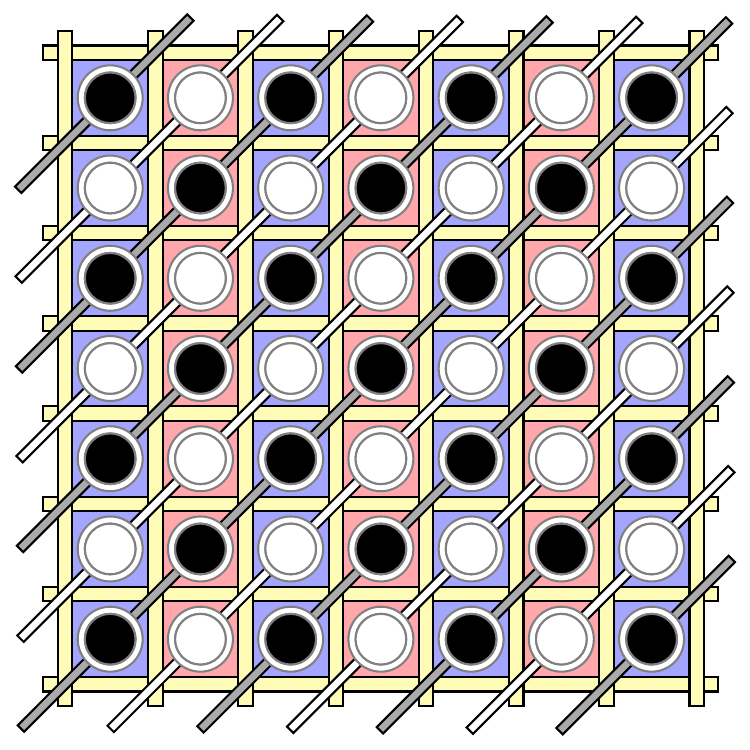}
    \caption{Schematics of the crossbar architecture. Circles represent quantum dots, separated by barrier gates (yellow), with on-site energies controlled by the diagonal plunger gates (gray and white lines). 
    Black circles are the electrons (or holes, in p-type devices); the spin of each electron is a qubit. 
    In the odd columns (light blue background) each qubit has Larmor frequency $\omega_o$, whereas in the even columns (light red background) each qubit has Larmor frequency $\omega_e$.}
    \label{fig:crossbar-setup}
\end{figure}

The crossbar architecture proposed and studied in \cite{Li_2018,Helsen_2018} consists of a two-dimensional array of quantum dots (see Fig.~\ref{fig:crossbar-setup}) with barrier gates (yellow lines in Fig.~\ref{fig:crossbar-setup}) separating them horizontally and vertically as well. These barrier gates control the interdot tunneling amplitudes, whereas the dot potentials (\emph{on-site energies}) are controlled via diagonal control lines (\emph{plunger gates}, white and gray in Fig.~\ref{fig:crossbar-setup}) which in turn enables the tuning of the charge configuration. A recent experiment has already demonstrated shared control in a $4\times 4$ crossbar array in planar germanium \cite{Borsoi_2023}.

To execute a quantum circuit on quantum computing hardware, the circuit has to be translated to a physical pulse sequence. 
In the case of this spin qubit crossbar architecture, the physical pulse sequence consists of gate-voltage pulses that govern the spatial motion of the electrons, and microwave magnetic pulses that trigger single-qubit logical gates.

\begin{figure}
    \centering
    \includegraphics[width=0.36\textwidth]{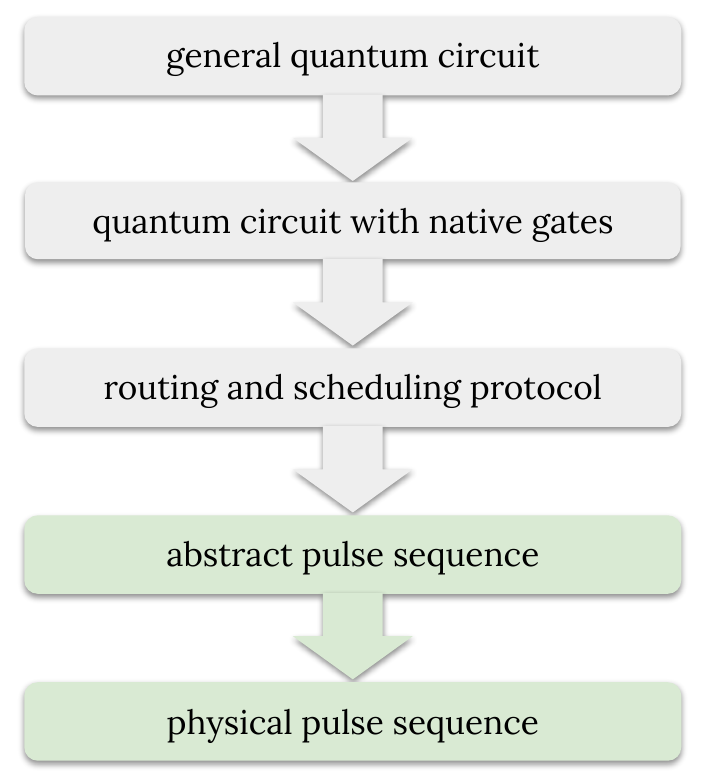}
    \caption{Layers for the quantum computing stack for a crossbar spin-qubit architecture.
    Special layers, characteristic of this architecture, are the `routing and scheduling protocol', which describes the motion of the electrons in the array, and the `abstract pulse sequence', which encodes the structure of subsequent gate-voltage settings required to move the electrons according to the routing and scheduling protocol.  
    See text for a more detailed description of all layers.
    Layers denoted by gray color were also studied in earlier works: for the surface code in Ref.~\cite{Helsen_2018} and in a general compilation framework in Ref~\cite{Paraskevopoulos_2023}. Layers denoted by green color are addressed in this work.}
    \label{fig:control-stack-layers}
\end{figure}

We envision this translation procedure as a sequence of layers, which we depict in Fig.~\ref{fig:control-stack-layers}.
Here, we describe these layers briefly, and leave details and examples for the bulk of the paper. 
The quantum circuit serves as the input for this procedure. 
The first step is to express the quantum circuit in terms of the native gates of the architecture.
The second step, characteristic of sparse spin qubit arrays (as well as other architectures where the qubits are mobile, e.g., trapped-ion and neutral-atom quantum computers), is to design the motion of the qubits on the lattice; this layer is termed \emph{routing and scheduling protocol} in Fig.~\ref{fig:control-stack-layers}.
A subsequent task is to translate the protocol to an \emph{abstract pulse sequence}: most importantly, this encodes the structure of subsequent gate-voltage settings that are required to move the electrons as specified by the routing and scheduling protocol.
Finally, the physical pulse sequence provides a physical specification of the abstract pulse sequence, e.g., expressing the gate-voltage settings in microvolts (or microelectronvolts) and timings in microseconds, shapes of magnetic-field pulses specified in milliteslas (or microelectronvolts) and microseconds, etc.

After the proposal \cite{Li_2018}, follow-up work devised a routing and scheduling protocol for the surface code implemented in the crossbar architecture \cite{Helsen_2018}, and proposed quantum circuit compilation strategies for the crossbar architecture \cite{Paraskevopoulos_2023,paraskevopoulos2024besnake}, down to the `routing and scheduling protocol' layer of the stack (see Fig.~\ref{fig:control-stack-layers}).
Our results are complementary to these earlier works, as we focus on the remaining layers of the stack, the `abstract pulse sequence' and `physical pulse sequence' layers (denoted green in Fig.~\ref{fig:control-stack-layers}).

In this work, we study the opportunity of realizing the surface code on the crossbar architecture studied in Refs.~\cite{Li_2018,Helsen_2018}.
Our contributions are enumerated as follows:
\begin{enumerate}[(a)]
  \item We decompose the surface code stabilizer measurement circuits in terms of single-qubit and two-qubit gates that are native to the spin-qubit architecture. This is a correction of the decomposition in Ref.~\cite{Helsen_2018}.
  \item We describe an updated routing and scheduling protocol (compared to Ref.~\cite{Helsen_2018}) for the implementation of such a stabilizer measurement cycle.
  \item We make an estimate for the scaling of logical error with code distance, based on the framework established in Sec.~5.4 of Ref.~\cite{Helsen_2018}, but applying that to our updated protocol.
  \item The surface code in this crossbar architecture can be operated in a \emph{line-by-line} or in a \emph{parallel} fashion \cite{Helsen_2018}. 
  The logical error probability, and its scaling with code distance $d$, depends strongly on this choice.  
  For line-by-line operation, we determine the optimal code distance as a function of the physical qubit coherence time and the measurement time, and identify a parameter range where the logical error rate of the surface code quantum memory is less than the physical error rate (often referred to as the \emph{break-even} performance of the code). 
  \item We construct an abstract pulse sequence (i.e., for gate voltage and ESR pulses) for the parallel operation of the surface code stabilizer measurement cycle.
  \item We devise a verification algorithm for adiabatic qubit shuttling-based routing in the shared-control crossbar architecture. 
  As inputs, one needs to specify (i) the desired movements of the electrons (qubits) in the array, and (ii) the abstract pulse sequence specifying the gate-voltage changes. Then, the algorithm verifies if the abstract pulse sequence indeed generates the desired movements.
  \item We describe how to implement the abstract pulse sequence physically.
  \item During the physical pulse sequence, crosstalk effects (namely, coherent phase errors) can occur on idle qubits, due to the built-in constraints of the crossbar architecture. 
  We characterize these crosstalk errors during the surface code cycle.  We also identify an experimentally relevant parameter regime where the estimated crosstalk errors are below the surface code threshold. These results suggest that these crosstalk errors can be mitigated by the surface code quantum error correction procedure. 
\end{enumerate}

As seen in the above list, contributions (a), (b), (c), and (d) are corrections or modifications or slight extensions with respect to \cite{Helsen_2018}.
The errors we have identified in \cite{Helsen_2018}, along with our proposed corrections, are summarized in Table~\ref{tab:list-of-corrections} in Appendix~\ref{appendix:line-by-line}.
We note that the errors we have identified are quantitative in nature, not affecting the qualitative conclusions of \cite{Helsen_2018}.
Our contributions (e), (f), (g), (h) go beyond the scope of \cite{Helsen_2018}.

The rest of the paper is organized as follows. 
In Sec.~\ref{sec:crossbar-intro}, we describe the key features of the crossbar spin qubit architecture.
In Sec.~\ref{sec:surface-code-circuits-and-protocol} we summarize the quantum circuits needed for the surface code stabilizer measurement cycle, we decompose them using the native spin qubit gates and we describe the updated scheduling protocol to transpile these circuits to the crossbar architecture (a--b). Moreover, we estimate the scaling of the logical error probability using the framework of \cite{Helsen_2018} (c--d). 
In Sec.~\ref{sec:pulse-sequence} we propose a physical pulse sequence for the stabilizer measurement cycle, and compute the crosstalk errors that occur during our protocol (e--h).
We provide a discussion of our results in Sec.~\ref{sec:discussion}, and conclude in Sec.~\ref{sec:conclusions}.

\section{The crossbar spin qubit architecture} \label{sec:crossbar-intro}

In this section, we briefly introduce the crossbar spin qubit architecture \cite{Li_2018,Helsen_2018}, which is shown in Fig.~\ref{fig:crossbar-setup}.

It consists of a two-dimensional array of quantum dots with barrier gates (yellow bars in Fig.~\ref{fig:crossbar-setup}) separating the dots horizontally and vertically. These barrier gates control the interdot tunneling amplitudes. 
The dot potentials (\emph{on-site energies}) are controlled via diagonal control lines (a.k.a. \emph{plunger gates}, shown as white and gray bars in Fig.~\ref{fig:crossbar-setup}) which enable the tuning of the charge configuration (up to the constraints originating from the shared-control structure).

Following Refs.~\cite{Li_2018,Helsen_2018}, we consider Loss--DiVincenzo qubits \cite{Loss_1998}, i.e. single-electron (or single-hole) spin qubits. A static external magnetic field is applied throughout the array to set the qubit splitting. However, in the proposal of the crossbar architecture, there is a built-in inhomogeneity: the qubit Larmor frequencies are alternating column by column \cite{Li_2018}, as indicated by the blue and red shadings in Fig.~\ref{fig:crossbar-setup}. This alternating structure enables selective (semi-global) addressing of the qubits located at even or odd columns, and single-qubit gates can be performed via Electron Spin Resonance (ESR), applying an a.c. magnetic field that has a transverse component with respect to the static magnetic field. 

Qubits can also move on the grid to unoccupied sites via coherent electron shuttling \cite{Fujita_2017,Mills_2019,Ginzel_2020,Krzywda_2020,Krzywda_2021,Jadot_2021,Yoneda_2021,Buonacorsi_2020,Seidler_2022,Zwerver_2023,van_Riggelen_Doelman_2024,Xue_2024}.
Moving the qubits between two sites with different Larmor frequencies, $\omega_e$ and $\omega_o$ (for brevity, we refer to angular frequencies as frequency throughout the paper), can effectively induce a coherent rotation around the axis of the external magnetic field.
Such a rotation can serve as a single-qubit gate \cite{Wang_2024}. 

A native two-qubit gate of spin qubit architectures is the $\sqrt{\text{SWAP}}$ gate that is carried out utilizing the exchange interaction between two neighboring qubits \cite{Loss_1998,JPetta_Science_2005,Sigillito_2019}. In the crossbar architecture, this is realized by moving the qubits to neighboring sites and lowering the potential barrier that separates them \cite{Noiri_2022_shuttling}. 
The diagonal plunger gates could also be used for radiofrequency (RF) charge sensing via gate reflectometry, enabling qubit readout via capacitance measurement at the Pauli spin blockade (PSB) transition \cite{Li_2018}. 
Qubit readout is also possible via charge sensors located in the vicinity of the quantum dots of the array \cite{Borsoi_2023}.

Importantly, all these considerations rely on a high level of uniformity throughout the spin qubit processor. 
Throughout Sec.~\ref{sec:surface-code-circuits-and-protocol} and \ref{sec:pulse-sequence}, we assume perfect uniformity of the array.
Also note that due to the shared control, crosstalk effects (e.g. unwanted rotations, spurious shuttlings) might also be induced, as we describe in Sec.~\ref{sec:idle-errors}.

\section{Surface code stabilizer measurement cycles} \label{sec:surface-code-circuits-and-protocol}

\begin{figure}
    \centering
    \includegraphics[width=0.47\textwidth]{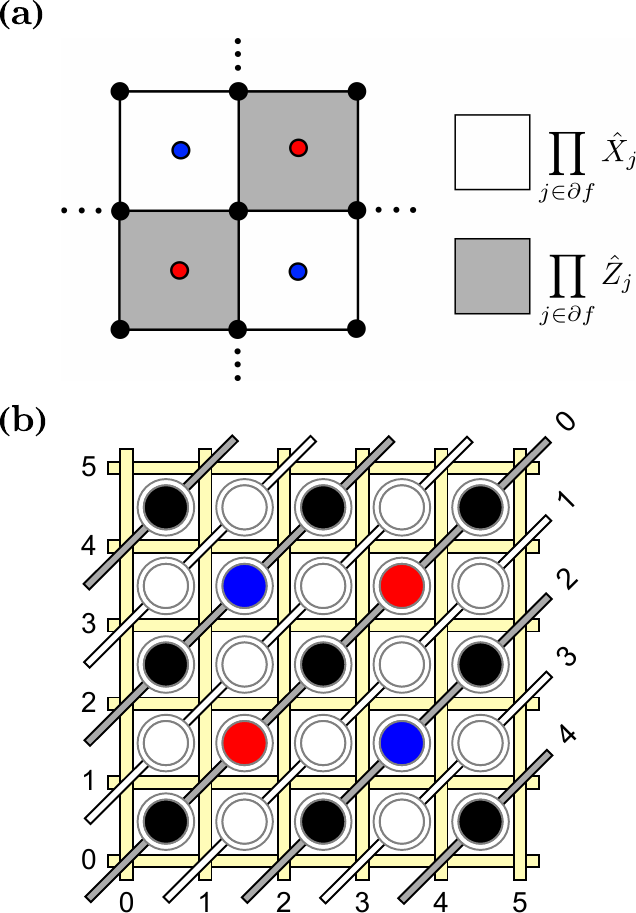}
    \caption{(a) Layout of the bulk of a surface code patch. The black dots are data qubits. Gray (white) plaquettes represent Z-type (X-type) stabilizer operators. Correspondingly, the red (blue) dots are the ancilla qubits for the Z (X) parity checks. We also use the notation ancilla qubit A and B for the red and blue qubits, respectively. (b) Same setup realized in a $5\times 5$ crossbar array.}
\label{fig:surface-code-setup}
\end{figure}

The surface code encodes a logical qubit in the common $+1$ eigenspace of the X-type and Z-type parity check operators (stabilizers), as shown in Fig.~\ref{fig:surface-code-setup}(a). In this section, we briefly summarize the quantum circuits of stabilizer measurements and their decomposition with native gates of the crossbar spin qubit architecture. Such a decomposition was included in Ref.~\cite{Helsen_2018}, but we identify a few errors there, which we correct. 
These errors and the proposed corrections are collected in Table~\ref{tab:list-of-corrections} in Appendix~\ref{appendix:line-by-line}. 

This section is structured as follows.
In Sec.~\ref{sec:surface-code-cycle-with-CNOTS} we list the quantum circuits for the surface code stabilizer measurement cycle \cite{Fowler_2012}.
In Sec.~\ref{sec:surface-code-cycle-with-sqrt-SWAPs} we decompose these circuits in terms of native gates of the crossbar spin qubit architecture.
In Sec.~\ref{sec:parallel-protocol} we provide an updated version of the scheduling protocol described in Sec.~4.4 of Ref.~\cite{Helsen_2018}, using parallel operation.
In Sec.~\ref{sec:line-by-line} we discuss the case when parallel operation is not feasible.
Finally, in Sec.~\ref{sec:error-scaling} we estimate the scaling of the logical error probability for a crossbar system, using the phenomenological error model introduced in Ref.~\cite{Helsen_2018}.

\subsection{Z-cycle circuit, X-cycle circuit with CNOT gates} \label{sec:surface-code-cycle-with-CNOTS}

\begin{figure}
    \centering
    \includegraphics[width=0.32\textwidth]{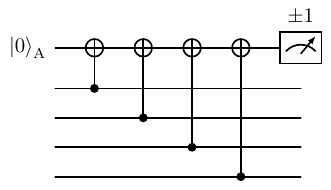}
    \caption{Surface code Z-stabilizer measurement circuit \cite{Fowler_2012}.}
    \label{fig:z-cycle-with-cnots}
\end{figure}

\begin{figure}
    \centering
    \includegraphics[width=0.38\textwidth]{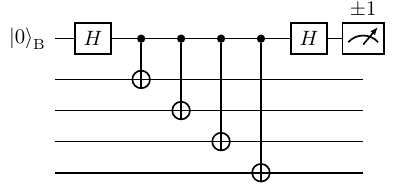}
    \caption{Surface code X-stabilizer measurement circuit \cite{Fowler_2012}.}
    \label{fig:x-cycle-with-cnots}
\end{figure}

Experimentally, the measurement of the stabilizers, i.e., checking the joint parity of data qubits on a plaquette, is performed by measuring an ancillary qubit after entangling the latter with the four corresponding data qubits. 
In principle, this is possible using one ancilla per lattice face [see Fig.~\ref{fig:surface-code-setup}(a)] exploiting the standard quantum circuits depicted in Fig.~\ref{fig:z-cycle-with-cnots} and Fig.~\ref{fig:x-cycle-with-cnots} for the Z- and X-type stabilizers, respectively \cite{Fowler_2012}. Both circuits use four CNOT gates for entangling data qubits with the ancilla. Ancilla qubits marked red in Fig.~\ref{fig:surface-code-setup}(a) are labeled by A in Fig.~\ref{fig:z-cycle-with-cnots}, and they are used to measure the Z-type stabilizers.
Similarly, ancilla qubits marked blue in Fig.~\ref{fig:surface-code-setup}(a) are labeled by B in Fig.~\ref{fig:x-cycle-with-cnots} and they are used to measure the X-type stabilizers.
For the X-type stabilizer measurement, two Hadamard gates are also needed, as shown in Fig.~\ref{fig:x-cycle-with-cnots}. During a complete Z-cycle (X-cycle), all of the Z (X) stabilizers are measured.

\subsection{Z-cycle circuit, X-cycle circuit using $\sqrt{\text{SWAP}}$ gates} \label{sec:surface-code-cycle-with-sqrt-SWAPs}

\begin{figure}
    \centering
    \includegraphics[width=0.48\textwidth]{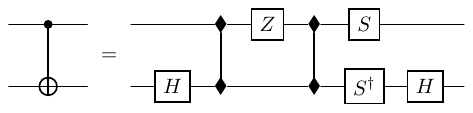}
    \caption{Decomposition of the CNOT gate in terms of $\sqrt{\text{SWAP}}$ gates and single-qubit gates that are native to the crossbar spin qubit architecture.}
    \label{fig:cnot-with-sqrt-swaps}
\end{figure}

Here, following \cite{Helsen_2018}, we compile the circuits of Fig.~\ref{fig:z-cycle-with-cnots} and Fig.~\ref{fig:x-cycle-with-cnots} such that they utilize the $\sqrt{\text{SWAP}}$ gate \cite{Loss_1998} instead of the CNOT as the standard two-qubit gate.
The stabilizer measurement circuits can be decomposed using the identity shown in Fig.~\ref{fig:cnot-with-sqrt-swaps} which uses two $\sqrt{\text{SWAP}}$ gates and five single-qubit gates for the CNOT \cite{Loss_1998}. For the single-qubit gates, we use Hadamard gates, the Pauli-$Z$ gate, the $S$ gate,
\begin{equation}
    S = \begin{pmatrix}
        1 & 0 \\
        0 & i 
        \end{pmatrix},
\end{equation}
and its Hermitian conjugate, the $S^\dag$ gate.

\begin{figure*}
    \centering
    \includegraphics[width=0.95\textwidth]{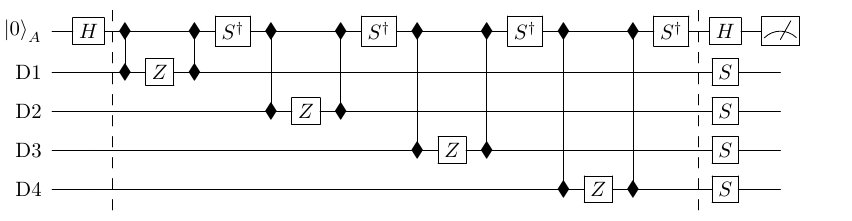}
    \caption{Surface code Z-stabilizer measurement circuit decomposed with two-qubit $\sqrt{\text{SWAP}}$ gates and single-qubit gates. In our Protocol, the gates before the first and after the second dashed line are realized by global ESR control. The gates between the first and second dashed lines are realized by shuttling. Labeling of the data qubits (D1, D2, D3, and D4) is consistent with Fig.~\ref{fig:unit-cell}.}
    \label{fig:z-cycle-with-sqrt-swaps}
\end{figure*}

\begin{figure*}
    \centering
    \includegraphics[width=0.95\textwidth]{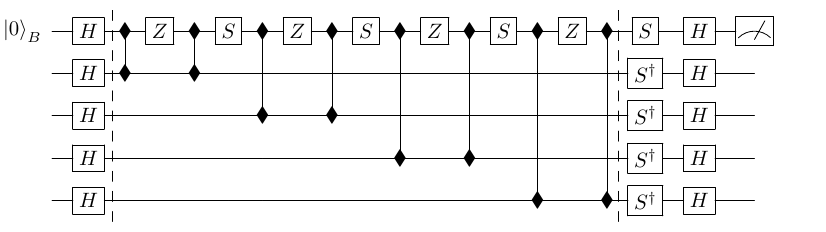}
    \caption{Surface code X-stabilizer measurement circuit decomposed with two-qubit $\sqrt{\text{SWAP}}$ gates and single-qubit gates. In our Protocol, the gates before the first and after the second dashed line are realized by global ESR control. The gates between the first and second dashed lines are realized by shuttling.}
    \label{fig:x-cycle-with-sqrt-swaps}
\end{figure*}

Substituting this decomposition into the previous circuits of Fig.~\ref{fig:z-cycle-with-cnots} and Fig.~\ref{fig:x-cycle-with-cnots}, after merging the subsequent Hadamard gates, we obtain the Z and X stabilizer measurement circuits with native gates of the crossbar architecture as shown in Fig.~\ref{fig:z-cycle-with-sqrt-swaps} and Fig.~\ref{fig:x-cycle-with-sqrt-swaps}, respectively. Note also that we shifted the mid-circuit $S$ and $S^\dag$ gates to the end of the circuits wherever it was possible.

\subsection{Routing and scheduling protocol for parallel operation} \label{sec:parallel-protocol}

In this section, we consider a protocol where the Z and X stabilizer measurement cycles are performed sequentially. Readout via PSB requires two quantum dots, therefore we are swapping the role of X and Z ancilla qubits in the subsequent cycles, using one of them always as a reference qubit for the measurement of the other.
We discuss how to implement (compile) the corresponding circuits to the crossbar architecture overcoming the hardware constraints.

Fig.~\ref{fig:surface-code-setup}(b) shows a $5\times 5$ crossbar array hosting $13$ qubits in the bulk of a surface code patch. Data qubits are black, Z-type ancilla qubits (ancilla qubits A) are red, and X-type ancilla qubits (ancilla qubits B) are blue.

For the surface code Z-cycle, using the quantum circuit of Fig.~\ref{fig:z-cycle-with-sqrt-swaps}, we propose an updated version of the protocol outlined in Sec.~4.4 of Ref.~\cite{Helsen_2018}.
It incorporates corrections 1 and 2 of Table~\ref{tab:list-of-corrections}, moreover, it also contains new steps (Steps~\ref{step:ancilla-H} and \ref{step:ancilla-H-2}). The short description of this Protocol is the following (for a visual representation see auxiliary file ZcycleMovie.pdf available at the Zenodo repository~\cite{suppmat_Zenodo}):
\begin{enumerate}
    \item Initialize the idle configuration. \label{step:init}
    \item Apply global $H$ gate to all ancilla qubits (located in even columns). \label{step:global-H}
    \item Go to the `rightward triangle configuration', by moving the ancilla qubits one site to the right, see Fig.~7(b) of \cite{Helsen_2018}. \label{step:right-triangle}
    \item Apply global $H$ gate to ancilla qubits B (located in even columns), to cancel the rotation they suffered in Step~\ref{step:global-H}. \label{step:ancilla-H}
    \item Perform the $\sqrt{\text{SWAP}}$ gates that entangle ancilla qubits A and the neighboring two data qubits (e.g. the gates that entangle ancilla qubit A and data qubits D1 and D2 in Fig.~\ref{fig:z-cycle-with-sqrt-swaps}), interleaved with shuttling-based single-qubit rotations $Z$ and $S^\dag$ on the data qubits and the ancilla qubit, respectively. \label{step:sqrt-of-swaps}
    \item Go back to idle configuration. \label{step:back-to-idle}
    \item Go to `leftward triangle configuration', by moving the ancilla qubits one site to the left \cite{Helsen_2018}. \label{step:left-triangle}
    \item As a mirror image of Step~\ref{step:sqrt-of-swaps}, perform the two $\sqrt{\text{SWAP}}$ gates that entangle ancilla qubits A with the other two neighboring data qubits (e.g. the gates that entangle ancilla qubit A and data qubits D3 and D4 in Fig.~\ref{fig:z-cycle-with-sqrt-swaps}), interleaved with shuttling-based single-qubit rotations $Z$ and $S^\dag$ on the data qubits and the ancilla qubit, respectively. \label{step:sqrt-of-swaps-2}
    \item Apply global $H$ gate to ancilla qubits B (located in even columns). \label{step:ancilla-H-2}
    \item Go back to idle configuration. \label{step:back-to-idle-2}
    \item Apply global $H$ gate to all ancilla qubits (located in even columns) and global $S$ gate to all data qubits (located in odd columns). \label{step:global-H-2}
    \item Go to 'rightward triangle configuration' \cite{Helsen_2018}. \label{step:measurement-conf}
    \item Perform PSB readout of ancilla qubits A using ancilla qubits B as reference qubits. \label{step:readout}
    \item Go back to idle configuration. \label{step:measurement-to-idle-conf}
\end{enumerate}

We propose to perform the mid-circuit single-qubit gates (i.e., those between the vertical dashed lines in Fig.~\ref{fig:z-cycle-with-sqrt-swaps} and \ref{fig:x-cycle-with-sqrt-swaps}) via coherent qubit shuttling since all of them are rotations around the z-axis. When the external magnetic field is pointing in the z-direction, this is feasible with the proper timing. Single-qubit gates before the first and after the second vertical dashed line in Fig.~\ref{fig:z-cycle-with-sqrt-swaps} and \ref{fig:x-cycle-with-sqrt-swaps} are conveniently performed by global rotations using ESR \cite{Li_2018,Helsen_2018}, because the same gate has to be applied to all qubits located in all columns with the same color. 
ESR enables global rotations around any axis (e.g. with the combination of $x$ and $y$ axis rotations),
thus $S$ and  $S^\dag$ gates can also be realized in this manner, as well as the Hadamard gate $H$. Note that compared to Fig.~\ref{fig:z-cycle-with-sqrt-swaps}, the circuit corresponding to our routing and scheduling Protocol contains two extra Hadamard gates (Steps~\ref{step:ancilla-H} and \ref{step:ancilla-H-2}) on ancilla qubits~B.

For the surface code X-cycle, we propose a similar protocol exchanging the role of ancilla qubits A with ancilla qubits B, based on the quantum circuit displayed in Fig.~\ref{fig:x-cycle-with-sqrt-swaps}. 

\begin{figure}
    \centering
    \includegraphics[width=0.47\textwidth]{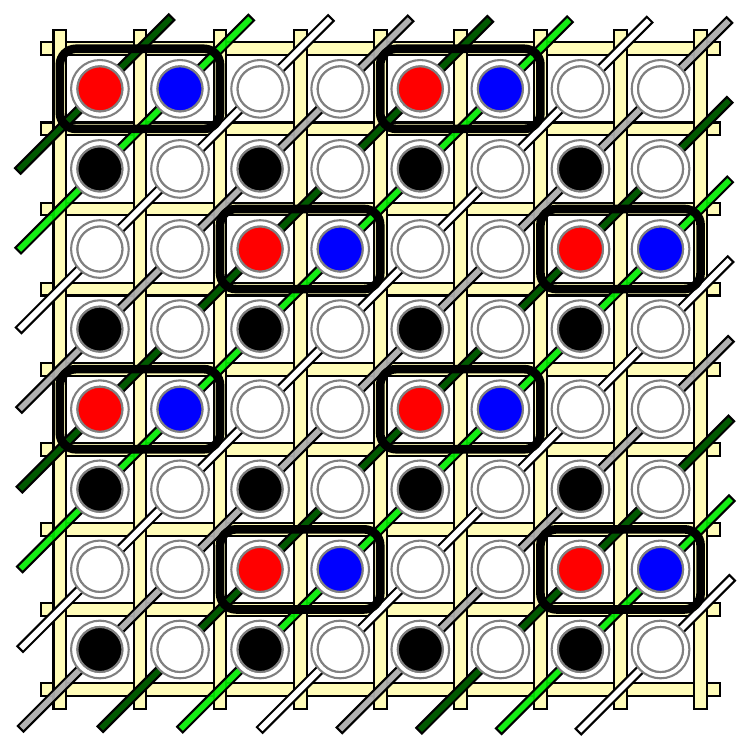}
    \caption{Parallel Pauli spin blockade readout of the red qubits (ancilla qubits A), using the blue qubits (ancilla qubits B) as reference qubits, in the surface-code Z-cycle. Charge sensors (not shown) are needed to perform readout in this configuration; these could be a set of sensor dots on the same chip \cite{Borsoi_2023}, or sensors added in a 3D fashion \cite{PettaTip,Futaya_2024,Rosenberg_2017,Conner_2021}.
    The crossbar array is in the rightward triangle configuration~\cite{Helsen_2018}. 
    On-site energies beneath the dark green, light green, gray, and white diagonal plunger gates have low to high values, respectively.
    The vertical barriers between the red and blue qubits are to be opened, and the light green diagonal plunger gates are to be far detuned to provide significantly lower on-site energy to overcome Coulomb repulsion between the red and blue qubits for the PSB readout.
    The same scheme can be used in the X-cycle, exchanging the role of the red and blue qubits.
    }
    \label{fig:parallel-measurement}
\end{figure}

In principle, in an X or Z-cycle, each step described above could be executed simultaneously in each $4\times 4$ unit cell of the crossbar array which has $4\times 4$ spatial periodicity (see Fig.~\ref{fig:unit-cell}). 
In an experimental setting, measurement of spin qubits is possible via spin-to-charge conversion and charge readout using charge sensors or gate-reflectometry \cite{Crippa_2019,ConnorsReflectometry}. The latter measures the charge response (i.e. the quantum capacitance) of tunnel-coupled qubits to changing plunger gate voltage that tunes the on-site energy difference. This measurement is possible via the diagonal plunger gates. 
Here, we investigate the case when charge sensors are used for the readout which has the advantage that vertical barrier activation is needed only for a short duration, reducing the amount of crosstalk errors (discussed in Sec.~\ref{sec:idle-errors}).
Utilizing multiple charge sensors and PSB-based spin-to-charge conversion, qubit states across the chip could be inferred simultaneously \cite{Borsoi_2023}. 
Therefore, we assume that ancilla qubits could be measured in parallel via the parallel PSB readout scheme depicted in Fig.~\ref{fig:parallel-measurement}.
Hence we term the protocol described in this section \emph{parallel operation}. 
In the next section, we discuss a scenario \cite{Helsen_2018} where operations are not executed in parallel across the array.

\subsection{Line-by-line implementation} \label{sec:line-by-line}

One could also opt for line-by-line operation throughout the whole cycle to account for qubit-to-qubit variations with individual control (see Sec.~'Parallel operation' of Ref.~\cite{Li_2018}, and Sec.~3 of Ref.~\cite{Helsen_2018}). 
\emph{Line-by-line operation} means that except for global rotations, each operation (quantum gate, qubit shuttling, measurement) is done sequentially, column-by-column, or row-by-row. In particular, after performing Steps~\ref{step:init} and \ref{step:global-H}, Steps~\ref{step:right-triangle}-\ref{step:back-to-idle} (excl. Step~\ref{step:ancilla-H} that is a global rotation) are applied only in the first two columns. At the end of this sequence, we return to the idle configuration, and then we continue with the next two columns, and so on, until reaching the end of the grid. This is followed by Steps~\ref{step:left-triangle}-\ref{step:measurement-conf} (excl. Step~\ref{step:ancilla-H-2}) performed in the same manner. 
Step~\ref{step:readout} is performed row-by-row.

Following Ref.~\cite{Helsen_2018}, we further divide each Step in our Protocol into \emph{composite time steps} so that one Step is equivalent to at least one composite time step. At this level, one composite time step corresponds to a qubit shuttling step or a gate in the quantum circuits of Fig.~\ref{fig:z-cycle-with-sqrt-swaps} and \ref{fig:x-cycle-with-sqrt-swaps}.
We use the adjective `composite' here, because below we will further divide these composite time steps to \emph{elementary time steps}.

In Appendix~\ref{appendix:line-by-line}, Table~\ref{tab:z-cycle-time-step-counts} lists the composite time-step count of each Step of our Protocol in terms of the types of operations for the line-by-line implementation of the Z parity checks. As a result of the line-by-line implementation, the number of composite time steps needed for most of the operations is proportional to the code distance $d$. This is not true for global rotations, which are applied in parallel for qubits located in the same column \cite{Helsen_2018}, taking just a single composite time step.
We get similar counts for the X stabilizer measurement cycle as that in Table~\ref{tab:z-cycle-time-step-counts}, the only difference is we need one more global rotation, and one less shuttling-based $z$ rotation for this protocol. Table~\ref{tab:total-time-step-count} summarizes the total number of composite time steps for all gate types for a full surface code stabilizer measurement cycle. Additionally, we list the number of operations per qubit type (data/ancilla qubit) for both cycles in Table~\ref{tab:gate-count}.
Similar considerations hold for parallel operation as well, substituting $d=1$ into the $d$-dependent entries of Table~\ref{tab:z-cycle-time-step-counts} and \ref{tab:total-time-step-count}.

\subsection{Dependence of the logical error on the code distance}
\label{sec:error-scaling}

In this section, we present the logical error probability estimation for a crossbar system based on Ref.~\cite{Helsen_2018}, using our parameters from the previous section. In Ref.~\cite{Fowler_2012}, surface code error correction was numerically simulated under realistic (circuit level) noise and the empirical formula, Eq.~(11) of Ref.~\cite{Fowler_2012},
\begin{equation} \label{eq:pL-fowler}
    p_L \approx 0.03\left(\dfrac{p_e}{8 p_{\text{th}}} \right)^{\frac{d+1}{2}} ,
\end{equation}
provided reasonable agreement with the simulation data for any odd distance $d$ (see Fig.~4 in Ref.~\cite{Fowler_2012}), where $p_L$ is the logical error probability per surface code cycle, $p_e$ is the per-cycle per-qubit physical error probability,
and $p_{\text{th}} = 0.57\%$ is the per-step threshold error probability. We will make use of this formula to estimate the error scaling for our Protocol.

In our estimate, we consider two types of errors: decoherence-induced errors and operation-induced errors. Decoherence is present in any realistic quantum system that is subject to an environment \cite{Nielsen_Chuang}. Therefore, even the dynamics of idling qubits are governed by decoherence processes. These are often separated into two categories: relaxation and dephasing. Both mechanisms are characterized by a corresponding time scale, $T_1$ and $T_2^\ast$ respectively.
For spin qubits, usually dephasing is the dominant error mechanism, meaning $T_2^\ast \ll T_1$ \cite{Burkard_2023}. Thus, for our simple estimate, we neglect relaxation processes; we consider only qubit dephasing as the source of decoherence. 

Assuming line-by-line operation (as discussed in Sec.~\ref{sec:line-by-line}), the total time of a full stabilizer measurement cycle using our Protocol is dependent on the code distance, and is calculated from Table \ref{tab:total-time-step-count} as follows:
\begin{equation} \label{eq:tau-tot-line-by-line}
    \tau_{\text{total}}^{\text{line-by-line}} (d) = 16 d \tau_{sw} + 15 d \tau_z + 10 d \tau_{sh} + 9\tau_{gl} + 2d \tau_m.
\end{equation}
Here, we used the data and notation in Table~\ref{tab:total-time-step-count}, and weighted the duration of each operation with the corresponding composite time-step count.
For parallel operation, the total time is independent of code distance:
\begin{equation} \label{eq:tau-tot-parallel}
    \tau_{\text{total}}^{\text{parallel}} = 16 \tau_{sw} + 15 \tau_z + 10 \tau_{sh} + 9\tau_{gl} + 2 \tau_m .
\end{equation}

Assuming an exponential decay of coherence, the decoherence-induced error probability is approximately given by the ratio of the total cycle time and the coherence time \cite{Helsen_2018}:
\begin{equation}
    P_{\text{decoh}}(d)  = 1 - e^{-\tau_{\text{total}} (d) / (2 T_2^\ast)} \approx \dfrac{\tau_{\text{total}} (d)}{2 T_2^\ast} .
\end{equation}
This is the standard way to introduce decoherence, however, we note that Gaussian decay - relevant for e.g. $1/f$ noise - could result in significantly better estimates for the logical error probability since for time scales shorter than $T_2^\ast$ it produces less decoherence.
For an optimistic estimate, we use $T_2^\ast = 1\unit{s}$, which is the measured Hahn-echo coherence lifetime of donor spin qubits in $^{28}$Si \cite{Li_2018,Tyryshkin_2011}. 
We will also use a less optimistic value, $T_2^\ast = 10\unit{ms}$ for comparison.

During quantum computation, single-qubit gates, two-qubit gates, shuttling, and measurements are also imperfect. To account for these imperfections we assign the same error probability, $10^{-3}$, to each operation, as listed in Table~\ref{tab:error-prob}. We also list the assumed time duration of the operations in Table~\ref{tab:error-prob}.
\begin{table}
    \centering
    \begin{tabular}{ccc}
    \hline\hline
        Operation                 &\,  Time duration &\, Error probability\\ \hline
        $\sqrt{\text{SWAP}}$ gate & $\tau_{sw} = 20\unit{ns}$ & $p_{sw} = 10^{-3}$ \\
        $Z$-rotation              & $\tau_{z} = 100\unit{ns}$ & $p_{z} = 10^{-3}$ \\
        Shuttling                 & $\tau_{sh} = 10\unit{ns}$ & $p_{sh} = 10^{-3}$ \\
        Global rotation           & $\tau_{gl} = 1000\unit{ns}$ & $p_{gl} = 10^{-3}$ \\
        Measurement               & $\tau_{m} = 100\unit{ns}$ & $p_{m} = 10^{-3}$ \\ \hline\hline
    \end{tabular}
    \caption{Operation times and error probabilities used for the estimation of the logical error probability $p_L$. All of the values are taken from Ref.~\cite{Helsen_2018}.}
    \label{tab:error-prob}
\end{table}

Taking into account both decoherence-induced and operation-induced errors, we obtain the average error probability per quantum stabilizer measurement cycle (per physical qubit):
\begin{align} \label{eq:P-tot-param}
\begin{split}
    P_{\text{avg}}(d) = 6 p_{sw} &+ 3.25 p_z + 3 p_{sh} \\
     & + 4.25 p_{gl} + 0.5 p_m + P_{\text{decoh}}(d).
\end{split}
\end{align}
Here, we also weighted the operation-induced error probabilities with the per-cycle average number of the specific operations (as listed in the last column of Table~\ref{tab:gate-count}). 
Qubits located at the code's boundary might have lower gate counts, which means that our estimate serves as an upper bound for the average error probability per cycle.

\begin{figure}
    \centering
    \includegraphics[width=0.48\textwidth]{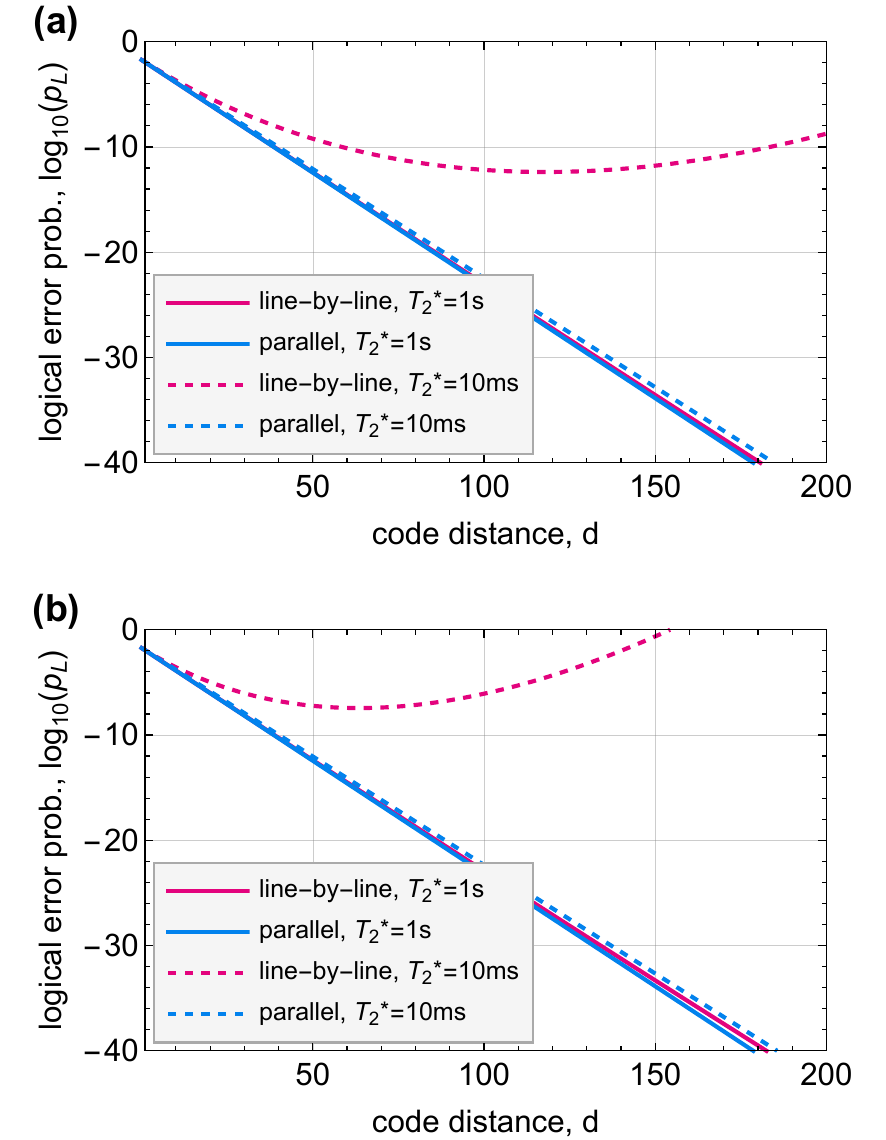}
    \caption{(a)~Logarithm of the estimated logical error probabilities based on Eq.~\eqref{eq:logical-error-rate}, as a function of code size, for measurement time $\tau_m = 100\unit{ns}$. Plotted for line-by-line and parallel operations, for coherence times $T_2^\ast = 1\unit{s}$ and $10\unit{ms}$ as well. Parallel operation implies error suppresion factors $\Lambda\approx 2.7$ and $2.6$ for coherence times $T_2^\ast = 1\unit{s}$ and $ 10\unit{ms}$, respectively. (b)~Same plot, but using $\tau_m = 1000\unit{ns}$ for the measurement time duration. 
    The error suppression is approximately the same for the two different dephasing time values in case of parallel operation.}
    \label{fig:logical-error-rate}
\end{figure}

To evaluate the logical error probability, now let us use Eq.~\eqref{eq:pL-fowler}, replacing the per-cycle error probability $p_e$ by our estimate $P_{\text{avg}}(d)$, yielding
\begin{equation}\label{eq:logical-error-rate}
    p_L = 0.03\left(\dfrac{P_{\text{avg}}(d)}{8 p_{\text{th}}} \right)^{\frac{d+1}{2}} .
\end{equation}
The thus obtained logical error probability is plotted in Fig.~\ref{fig:logical-error-rate} as a function of code distance, both for parallel and line-by-line operation, for two different values of $T_2^\ast$, using $p_{\text{th}} = 0.57\%$ \cite{Fowler_2012}. 
Panels (a) and (b) of Fig.~\ref{fig:logical-error-rate} correspond to two different values of the measurement time $\tau_\textrm{m}$.

In the case of parallel operation (blue solid and blue dashed lines in Fig.~\ref{fig:logical-error-rate}), the hardware constraints of the crossbar architecture do not imply any significant effect on the error suppression: the logical error probability goes below $p_L = 10^{-20}$ for code distance $d\approx 85$ for coherence times $T_2^\ast = 1\unit{s}$ and $T_2^\ast = 10\unit{ms}$ as well. This level of precision is suitable for a fully functioning computer.
Eq.~\eqref{eq:logical-error-rate} is often cast into the form
\begin{equation}\label{eq:logical-error-rate-with-Lambda}
        p_L = 0.03/ \Lambda^{\frac{d+1}{2}}
\end{equation}
as well, where $\Lambda = 8 p_{\text{th}} / P_{\text{avg}}$ is the exponential error suppression factor which is just a constant in case of parallel operation.
We extract the values $\Lambda\approx 2.7$ and $2.6$ for coherence times $T_2^\ast = 1\unit{s}$ and $10\unit{ms}$, respectively. These values remain approximately the same regardless of using $100\unit{ns}$ or $1000\unit{ns}$ measurement time.

However, for line-by-line operation, due to the linear scaling of the operation time with the code distance, the logical error probability has a global minimum at a certain code distance $d^\ast$ \cite{Helsen_2018}. 
For coherence time $T_2^\ast = 10\unit{ms}$ and measurement time $\tau_\textrm{m} = 100\unit{ns}$ (red dashed line in Fig.~\ref{fig:logical-error-rate}(a)), the minimum achievable logical error probability is around $p_L = 10^{-12}$ at code distance $d^\ast =117$. This becomes even less favorable when we increase the measurement time $\tau_m$ to $1000\unit{ns}$ (red dashed line in Fig.~\ref{fig:logical-error-rate}(b)), where the lowest feasible logical error probability is around $p_L = 10^{-7}$ at code distance $d^\ast =63$. 

\begin{figure*}
    \centering
    \includegraphics[width=0.99\textwidth]{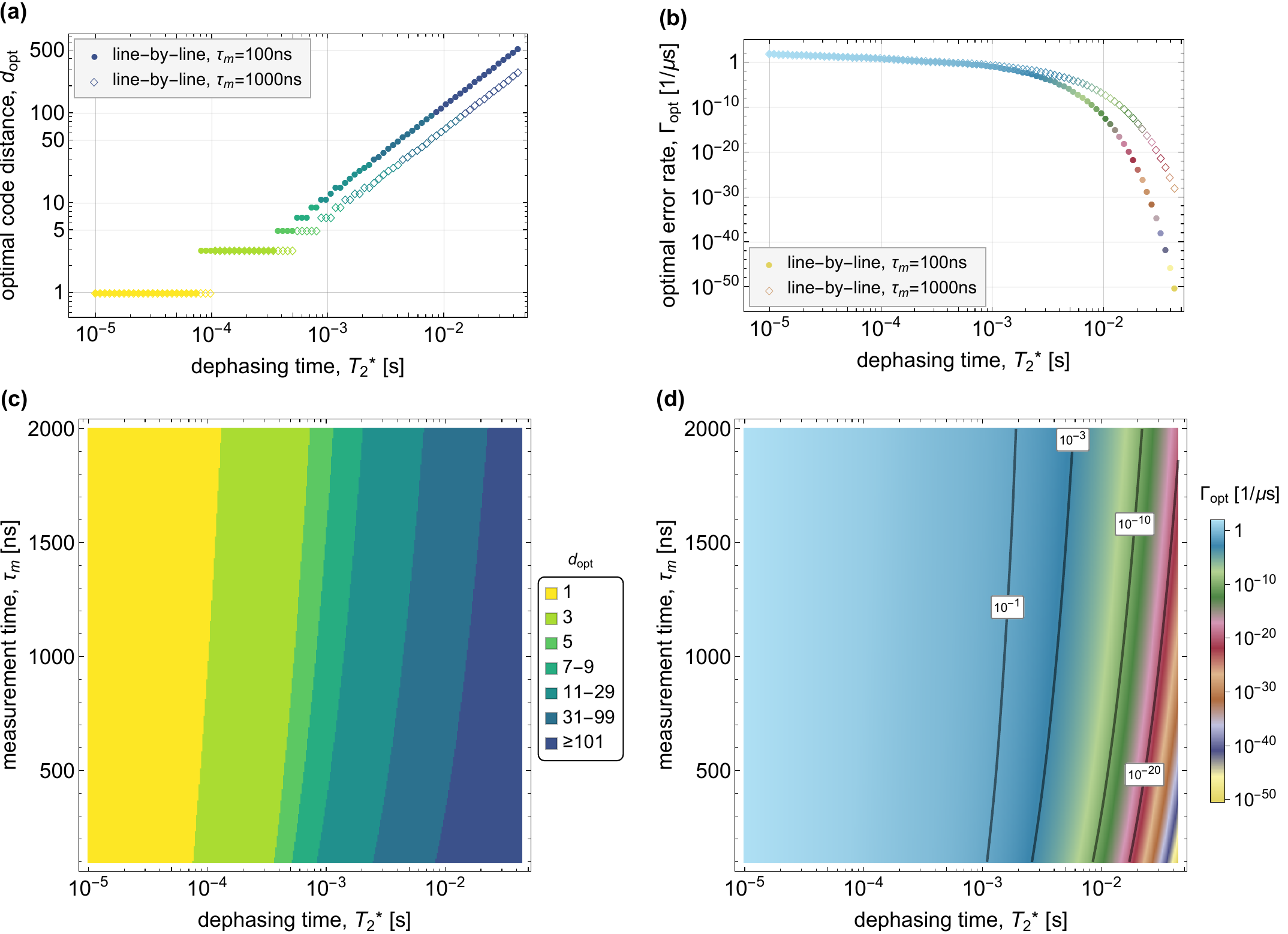}
    \caption{(a)~Optimal code distance of Eq.~\eqref{eq:optimal-distance}, for line-by-line operation as a function of the dephasing time $T_2^\ast$, for measurement time durations $\tau_m = 100\unit{ns}$ and $1000\unit{ns}$. (b)~The corresponding optimized error rates $\Gamma_{\text{opt}} = \Gamma (d_{\text{opt}})$ in units of $\unit{(\mu s)}^{-1}$. (c)~Optimal code distance shown as a function of both dephasing time and measurement time. (d)~The optimized error rate $\Gamma_{\text{opt}}$ shown as a function of both dephasing time and measurement time.}
    \label{fig:optimal-code-distance}
\end{figure*}

For the line-by-line implementation, we also compute the optimal code distance which we define as
\begin{equation} \label{eq:optimal-distance}
    d_{\text{opt}} = \arg\min_d \Gamma(d) ,
\end{equation}
where $\Gamma(d)$ is the error rate,
\begin{equation}
    \Gamma(d) = 
    \begin{cases} 
      1 / T_2^\ast & d = 1 \\
      p_L / \tau_{\text{total}}^{\text{line-by-line}} (d) & d \geq 3 .
   \end{cases}
\end{equation}
The optimal code distance as a function of the dephasing time is illustrated in Fig.~\ref{fig:optimal-code-distance}(a). 
Up until $T_2^\ast \approx 100\unit{\mu s}$ it is $1$, meaning that scaling up the surface code leads to increased logical error. We consider the surface code quantum memory useful when the encoded logical qubit has less error than a single physical qubit. Therefore, in Eq.~\eqref{eq:optimal-distance}, we compare the achievable logical error probability per cycle time to the error of a single idling physical qubit.
Fig.~\ref{fig:optimal-code-distance}(b) also shows the optimal (minimized) error rates $\Gamma (d)$ as a function of the dephasing time. 
Furthermore, we also plot the optimal code distance and the optimized error rate as a function of both the dephasing time and the measurement time (from $\tau_m = 100\unit{ns}$ to $2000\unit{ns}$) in Fig.~\ref{fig:optimal-code-distance}(c) and (d), respectively.
Fig.~\ref{fig:optimal-code-distance}(d) shows that above-millisecond coherence time is required to push the logical error rate per microsecond below $10^{-3}$, which is the error of a single operation.

\section{Pulse sequence for the stabilizer measurement cycles} \label{sec:pulse-sequence}

In Sec.~\ref{sec:parallel-protocol} we described a scheduling Protocol for the surface code stabilizer measurement cycle, although the question remains whether there is a practical pulse sequence that could realize it. In this section, we argue that the answer to this question is positive, furthermore, we construct a specific pulse sequence that is suitable for our Protocol. Following Fig.~\ref{fig:control-stack-layers}, in Sec.~\ref{sec:abstract-pulse-sequence}, first we describe this pulse sequence in terms of abstract gate voltage values and abstract ESR pulses. Based on this description, in Sec.~\ref{sec:shuttling-protocol}, we also provide a verification algorithm for the qubit shuttling steps included in our Protocol. We assign physical pulses to the abstract ones in Sec.~\ref{sec:physical-pulse-sequence}, establishing the final control stack layer of implementing the surface code with the crossbar architecture (Fig.~\ref{fig:control-stack-layers}).
Finally, in Sec.~\ref{sec:idle-errors}, we also estimate the idle qubit errors occurring during our implementation of the surface code stabilizer measurement cycle due to the crosstalk effects.

\subsection{Abstract pulse sequence} \label{sec:abstract-pulse-sequence}

Our Protocol described in Sec.~\ref{sec:parallel-protocol} consists of global single-qubit gates, qubit shuttling steps, shuttling-based single-qubit phase gates, two-qubit $\sqrt{\text{SWAP}}$ gates, and measurements. 
Coherent shuttling of a single qubit is a simple procedure, however, for parallel operation, it is a non-trivial problem to perform multiple shuttlings simultaneously, due to the limited control in the crossbar architecture. This is also true for the $\sqrt{\text{SWAP}}$ gates.

Assuming that throughout the quantum-dot array, the Coulomb energy and the orbital level spacing are much larger than the on-site energy differences, we propose that it is sufficient to use 4 different plunger gate voltage values to carry out our Protocol with parallel operation. 
(In fact, a fifth value is needed for the spin-to-charge conversion in Step~\ref{step:readout} of our Protocol.)
Furthermore, to describe our pulse sequence it is enough to prescribe the gate voltages on 4 consecutive diagonal plunger gates, and then we have a pattern that is periodically repeating itself.

At this level, we further divide composite time steps (used in Sec.~\ref{sec:line-by-line}) into \emph{elementary time steps} to resolve the pulsing of the plunger and barrier gates.
In Table~\ref{tab:abstract-pulse-sequence}, we outline the first 18 elementary time steps of the abstract gate voltage and abstract ESR pulse sequence corresponding to the surface code Z-cycle based on our Protocol.
Each column of this table corresponds to one elementary time step and contains a value (1, 2, 3, or 4) for the 4 consecutive diagonal plunger gates that is proportional to the value of the on-site energies, On/Off declarations for the barriers, and the ESR field as well. For the numbering of the plunger gates and the barriers, we use the same convention as in Fig.~\ref{fig:surface-code-setup}(b).

Starting from the idle configuration (elementary time step 1), elementary time step 2 is the Hadamard gate applied to the ancilla qubits, and elementary time steps 3-5 correspond to qubit shuttling to reach the rightward triangle configuration. Elementary time step 6 corresponds to Step~\ref{step:ancilla-H} in our scheduling Protocol, elementary time step 7 is a plunger reset, elementary time steps 8-10 correspond to the first $\sqrt{\text{SWAP}}$ of Step~\ref{step:sqrt-of-swaps}, elementary time steps 11-15 realize a shuttling based $Z$ gate, and elementary time steps 16-18 correspond to the second $\sqrt{\text{SWAP}}$ gate of Step~\ref{step:sqrt-of-swaps}.
Most of the Steps in our Protocol consist of more than one elementary time step, corresponding to several columns in Table~\ref{tab:abstract-pulse-sequence}, consequently.

At the Zenodo repository~\cite{suppmat_Zenodo}, we provide the full abstract gate-voltage pulse sequence for both the Z and X stabilizer measurement cycles implemented with parallel operation, in the form of two Tables (see auxiliary files ZcyclePulseSequence.xlsx and XcyclePulseSequence.xlsx, respectively).

\begingroup
\setlength{\tabcolsep}{5pt}
\begin{table*}
    \centering
    \begin{tabular}{cccccccccccccccccccc}
    \hline\hline
        Components & Specification & 1 & 2 & 3 & 4 & 5 & 6 & 7 & 8 & 9 & 10 & 11 & 12 & 13 & 14 & 15 & 16 & 17 & 18 \\ \hline
        \multirow{4}{*}{plunger gates} & 0 & 1 & 1 & 1 & c & 3 & 3 & 3 & 3 & 3 & 3 & 3 & c & 4 & c & 3 & 3 & 3 & 3 \\
                    & 1 & 4 & 4 & 4 & c & 2 & 2 & c & 3 & 3 & 3 & c & 2 & 2 & 2 & c & 3 & 3 & 3  \\
                    & 2 & 1 & 1 & 1 & 1 & 1 & 1 & 1 & 1 & 1 & 1 & 1 & 1 & 1 & 1 & 1 & 1 & 1 & 1  \\
                    & 3 & 4 & 4 & 4 & 4 & 4 & 4 & 4 & 4 & 4 & 4 & 4 & 4 & 4 & 4 & 4 & 4 & 4 & 4  \\[0.8em]
        \multirow{4}{*}{barrier gates}  & even, horizontal & 1 & 1 & 1 & 1 & 1 & 1 & 1 & c & 0 & c & 1 & 1 & 1 & 1 & 1 & c & 0 & c \\
                    & odd, horizontal & 1 & 1 & 1 & 1 & 1 & 1 & 1 & 1 & 1 & 1 & 1 & 1 & 1 & 1 & 1 & 1 & 1 & 1  \\
                    & even, vertical & 1 & 1 & c & 0 & c & 1 & 1 & 1 & 1 & 1 & 1 & c & 0 & c & 1 & 1 & 1 & 1 \\
                    & odd, vertical & 1 & 1 & 1 & 1 & 1 & 1 & 1 & 1 & 1 & 1 & 1 & 1 & 1 & 1 & 1 & 1 & 1 & 1 \\[0.8em]
        \multirow{2}{*}{ESR}  & $\omega_o$ & 0 & 0 & 0 & 0 & 0 & 0 & 0 & 0 & 0 & 0 & 0 & 0 & 0 & 0 & 0 & 0 & 0 & 0  \\
                    & $\omega_e$ & 0 & 1 & 0 & 0 & 0 & 1 & 0 & 0 & 0 & 0 & 0 & 0 & 0 & 0 & 0 & 0 & 0 & 0  \\
        \hline\hline
    \end{tabular}
    \caption{First 18 elementary time steps of the (abstract) gate voltage and ESR pulse sequence realizing the surface code Z-cycle. For the numbering of the plunger gates and the barriers, we use the same convention as in Fig.~\ref{fig:crossbar-setup} and Fig.~\ref{fig:surface-code-setup}(b). Plunger gate voltages are periodic before and after the four we listed in this table. Here, we associate integers with the plunger voltage values. A higher integer value corresponds to a higher gate voltage (on-site potential). For the barrier gates, we have On/Off declarations corresponding to 1 or 0, respectively. The value "c" refers to that the gate voltage is changing. For the ESR pulses, 0 means no pulse and 1 means a Hadamard gate. The full sequence (111 elementary time steps) together with a visual representation is provided at the Zenodo repository~\cite{suppmat_Zenodo}, see ZcyclePulseSequence.xlsx and ZcycleMovie.pdf correspondingly.}
    \label{tab:abstract-pulse-sequence}
\end{table*}
\endgroup

\subsection{Verification of the shuttling protocol} \label{sec:shuttling-protocol}

In this section, we summarize how we verify our shuttling protocol. 
Verification means two things:
we ensure that the (abstract) gate voltage pulse sequence meets the operational constraints of the crossbar architecture, and we check whether the given pulse sequence realizes the desired movements between the different spatial configurations of the electrons. Therefore, we use the following definitions and set the requirements listed below:

\begin{enumerate}
    \item Assume that each plunger gate voltage at each step of the pulse sequence takes one of the four different values: $1$, $2$, $3$, $4$.
    This abstract value represents the on-site energy controlled by the plunger gate, with $1$ ($4$) corresponding to the lowest (highest) on-site energy.
    \item An \textit{electron configuration} ($\mathtt{E}$) is the collection of electron positions in the lattice. Double occupancy of a single site is forbidden for shuttling.
    \item A \textit{plunger configuration} ($\mathtt{P}$) is a vector of plunger values.
    \item A \textit{configuration} $\mathtt{C} = (\mathtt{E},\mathtt{P})$ is an ordered pair of an electron configuration and a plunger configuration. \label{configuration}
    \item A plunger reset (also denoted by $\mathtt{P}$) describes the change of the plunger values. Formally, it is simply the plunger configuration after the change. If this is combined with barrier activation (see below), then it results in a change in the electron configuration. If it is not combined with barrier activation, then it does not change the electron configuration.
    \item A \textit{shift} ($\mathtt{S}$) describes the movement of electrons in the array in response to the activation of a barrier and a plunger reset. Formally, it is an ordered pair $\mathtt{S} = (\beta,\mathtt{P})$ where $\beta$ is the collection of barrier indices (of those barriers that are activated), and $\mathtt{P}$ is a plunger reset. Activation of the barriers means that first the barriers are lowered, then the plunger reset is applied, and then the barriers are raised back. \label{shift}
    \item We aim to avoid unwanted, and non-adiabatic tunneling processes, hence we constrain which shifts can be performed on which configurations. This is specified in the following:
    \begin{enumerate}
        \item Each shift must contain the activation of either horizontal or vertical barriers only.
        \item It is not allowed to activate the barriers on both sides of a dot in the same shift.
    \end{enumerate}
    \item A shift $\mathtt{S} = (\beta,\mathtt{P})$ is \textit{compatible} with a configuration $\mathtt{C}_0 = (\mathtt{E}_0, \mathtt{P}_0)$, if the following conditions are satisfied for all $b\in\beta$:
    \begin{enumerate}
        \item In the electron configuration $\mathtt{E}_0$, all neighboring dot pairs of barrier $b$ contains at most one electron.
        \item For all electrons in the neighboring dot pairs of barrier $b$, one side of the barrier has lower on-site energy than the other (also implying that the pre-shift plunger values specified in $\mathtt{P}_0$ are different), and the electron resides on the side of that barrier with the lower on-site energy.
        \item For all dot pairs neighboring barrier $b$, the post-shift plunger values specified in $\mathtt{P}$ are also different on the two dots.
    \end{enumerate}
    \item A shift is an operation that can act on configurations that are compatible with it. In such a case, the result (in other words, the \textit{action}) of the shift $\mathtt{S}=(\beta,\mathtt{P})$ on configuration $\mathtt{C}_0 = (\mathtt{E}_0, \mathtt{P}_0)$ is $\mathtt{C} = (\mathtt{E}, \mathtt{P})$, where $\mathtt{E}$ contains the new positions of the electrons. Only the positions of those electrons can change that initially reside in one of the dot pairs neighboring the barriers listed in $\beta$. Out of those electrons, the ones whose on-site energy remains lower after the shift will not move. Those electrons whose on-site energy is higher after the shift will move to the other dot of their dot pair.
    \item A \textit{shuttling protocol} is specified as an initial configuration, and a sequence of shifts and plunger resets (that are shifts without barrier activation), such that each shift is compatible with the configuration preceding it. Without the loss of generality, we can assume that bare plunger resets can be contracted, i.e., between two subsequent shifts there is at most one bare plunger reset. The result of a protocol is the final configuration $\mathtt{C}_N$, which can be expressed as the sequence of actions on the initial configuration:
    \begin{equation}
        \mathtt{C}_N = \mathtt{S}_{(N-1)\to N} \dots \mathtt{S}_{1\to 2} \mathtt{S}_{0\to 1} \mathtt{C}_0 ,
    \end{equation}
    where $N$ is the number of shifts in the shuttling protocol.
\end{enumerate}

Using this language, verification of a shuttling protocol is equivalent to checking whether all subsequent shifts are compatible with their prior configuration and whether their action indeed provides the desired configuration (for all of the intermediate steps as well).
If all of these requirements are satisfied, then the corresponding shuttling protocol is valid.

We implemented the above requirements as a verification algorithm in Python and tested the validity of the qubit shuttling steps of our Protocol (described in Sec.~\ref{sec:parallel-protocol})
by checking all the requirements for the $4\times 4$ unit cell of the grid (shown in Fig.~\ref{fig:unit-cell}), using periodic boundary conditions.
Our code is available at the Zenodo repository~\cite{Pataki_Verification_of_shuttling}.
It takes a list of configurations ($\mathtt{C}_0$, $\mathtt{C}_1$, $\mathtt{C}_2$,~\dots, $\mathtt{C}_N$) and shifts ($\mathtt{S}_{0\to 1}$, $\mathtt{S}_{1\to 2}$,~\dots, $\mathtt{S}_{(N-1)\to N}$) as input. 
As output, it returns boolean values depending on whether the performed shifts were compatible with the corresponding configurations and whether they resulted in the next configuration in the list. 
Since all shifts corresponding to shuttling steps (Step~\ref{step:right-triangle},~\ref{step:back-to-idle},~\ref{step:left-triangle},~\ref{step:back-to-idle-2},~\ref{step:measurement-conf}, and \ref{step:measurement-to-idle-conf}) in our Protocol were found to be compatible with their previous configuration and moved the grid to the subsequent configuration, we conclude that the relevant parts of our (abstract) gate voltage pulse sequence are valid.
We note that one can also use our verification algorithm to find shifts connecting two arbitrary configurations of the crossbar array by brute force checking all possible cases.

\subsection{Physical pulse sequence} \label{sec:physical-pulse-sequence}

In this section, we describe a physical implementation of the abstract gate voltage and ESR pulse sequence described in section \ref{sec:abstract-pulse-sequence}. 

As we mentioned in Sec.~\ref{sec:abstract-pulse-sequence}, it is sufficient to use 4 different plunger gate voltage values to carry out our Protocol. 
These plunger gate voltages control the quantum-dot on-site energies, to be used in the Hamiltonian introduced in Sec.~\ref{sec:idle-errors}.
Let us denote the quantum dot on-site energy values by $u_1$, $u_2$, $u_3$, and $u_4$, in increasing order. 
In what follows, we will change an on-site energy from $u_i$ to $u_j$ in time $T$ using the following smooth function:
\begin{equation} \label{eq:plunger-reset}
    \varepsilon(u_i,u_j,t) = u_i + \left( u_j - u_i \right) \sin^2\left( \dfrac{\pi t}{2T} \right).
\end{equation}
Similarly, we will tune a tunnel barrier separating neighboring quantum dots according to
\begin{equation} \label{eq:barrier-activation}
    t_0(t) = t_{0,\text{max}} \sin^2\left( \dfrac{\pi t}{T} \right),
\end{equation}
between the On ($t_0(T/2)=t_{0,\text{max}}$) and Off ($t_0(0)= t_0(T) = 0$) declarations.
We use Eq.~\eqref{eq:plunger-reset} to model the physical pulses corresponding to plunger resets, and Eq.~\eqref{eq:barrier-activation} to model barrier activation during our shuttling protocol (see Sec.~\ref{sec:shuttling-protocol}).

For shuttling-based $z$ rotations (see Step~\ref{step:sqrt-of-swaps} in the Protocol described in Sec.~\ref{sec:parallel-protocol}), we tune the plunger gate voltage of the target qubit ($u_\text{t}$) to the same level as the neighboring site ($u_\text{n}$) and we also tune it back; these two steps together take time $T_z$; in the same time window, we also activate the barrier: 
\begin{subequations} \label{eq:shuttling-based-z-rot}
\begin{align}
    \varepsilon_z (u_\text{t},u_\text{n},t) &= u_\text{t} + \left( u_\text{n} - u_\text{t} \right) \sin^2\left( \dfrac{\pi t}{T_z} \right), \\
    t_0(t) &= t_{0,\text{max}} \sin^2\left( \dfrac{\pi t}{T_z} \right).
\end{align}
\end{subequations}

The Hadamard gate can be expressed as a $\pi/2$ rotation around the $y$ axis, followed by a $\pi$ rotation around the $x$ axis,
\begin{equation} \label{eq:hadamard}
    H = X \sqrt{Y}.
\end{equation}
Similarly, the $S$ gate can be decomposed as
\begin{equation}
    S = \sqrt{X} \sqrt{Y} \sqrt{X}^\dagger.
\end{equation}
These elementary single-qubit rotations are executed by an ESR drive pulse
\begin{equation} \label{eq:ESR-drive-pulse}
    B_x(t) = B_{\text{ac}} \sin\left(\omega_{\text{ac}} t + \varphi\right) ,
\end{equation}
where $B_{\text{ac}}$ is the amplitude, $\omega_{\text{ac}}$ is the frequency (to be matched with $\omega_e$ or $\omega_o$, see Fig.~\ref{fig:crossbar-setup}) and $\varphi$ is the phase of the driving field. Specifically, a $\sqrt{Y}$ gate is obtained using $\varphi = \pi$ and pulse duration $\pi / (2\Omega)$, where $\Omega$ is the qubit Rabi frequency which is proportional to the drive strength $\Omega = \gamma B_{\text{ac}}/2$, where $\gamma$ is the electron gyromagnetic ratio. Similarly, a $\sqrt{X}$ gate is obtained using $\varphi = \pi /2$ and the same pulse duration. Doubled pulse durations provide the $Y$ and $X$ gates, respectively.
These gates are interpreted in the frame which is rotating with the qubit's Larmor frequency ($\omega_e$ or $\omega_o$) around the $z$ axis.

During $\sqrt{\text{SWAP}}$ gates, the two qubits involved are located at the same column, which means the Zeeman splitting is the same for both. In the crossbar architecture, these gates can be performed with a conventional exchange pulse \cite{Li_2018,Helsen_2018}, i.e., by lowering the barrier separating the two electrons such that the on-site energy of the two dots is kept equal.

Finally, we use the standard two-qubit double quantum dot Hamiltonian to describe the PSB readout \cite{Sen_2023}, assuming that the reference qubit is in $\ket{\uparrow}$ state occupying a quantum dot in an even column (with Larmor frequency $\omega_e$). The qubit to be measured (e.g. the red qubits in Fig.~\ref{fig:parallel-measurement}) is located on the neighboring site, in an odd column (with Larmor frequency $\omega_o$), and for concreteness, we assume that $\omega_o < \omega_e$. During the readout, the on-site energy of the reference qubit is lowered so that the detuning exceeds the Coulomb repulsion (we denote the corresponding value of plunger gate voltage by "R" in the abstract gate pulse sequence, see auxiliary files ZcyclePulseSequence.xlsx and XcyclePulseSequence.xlsx at the Zenodo repository~\cite{suppmat_Zenodo}), and the barrier between the qubits is opened simultaneously. Consequently, the anti-parallel spin configuration is converted to a single-site singlet; in contrast, the parallel spin configuration is blockaded and the two electrons remain separated on the two sites.
Finally, this charge difference is measured by a charge sensor and hence is used to infer the state state of the qubit.

\subsection{Idle-qubit errors due to limited control} \label{sec:idle-errors}

Due to the shared-control gate layout, crosstalk occurs: when an operation (e.g., shuttling, gate, readout) targets a subset of the qubits, then the remaining qubits, which are nominally idling, are also affected by the control fields, hence undergo undesired dynamics leading to errors.
We refer to this effect as crosstalk.
Note that our Protocol avoids opening a barrier between two idle qubits located at neighboring sites (see Sec.~\ref{sec:shuttling-protocol}), thus two-qubit crosstalk errors are absent.

In this section, we describe the crosstalk-induced, unwanted idle qubit rotations using the adiabatic approximation and an effective model. In Appendix~\ref{appendix:idle-qubit-errors}, we also provide an estimate for the rotation angles using experimentally relevant parameters.

We aim to characterize idle qubit errors during our Protocol, realized by the physical pulse sequence described in Sec.~\ref{sec:physical-pulse-sequence}.
Note that in our Protocol there is no such step or configuration when horizontal and vertical barriers are open simultaneously, meaning that horizontal and vertical movement of the qubits is completely separated during the stabilizer measurement cycle.
Whenever a vertical barrier is open, horizontal, tunnel-coupled double quantum dots are formed from the two columns neighboring that vertical barrier.

To model this scenario (e.g. a coherent shuttling) we use a minimal model: 
a single electron in a double quantum dot with Hamiltonian
\begin{align}\label{eq:HDQD}
    H_\text{DQD} &= H_{\text{os}} + H_\text{t} + H_\text{Z} , \\
	H_{\text{os}} &= \dfrac{\epsilon(t)}{2} \tau_z , \\
    H_\text{t} &= t_0(t) \tau_x , \\
	H_\text{Z} &= \dfrac{1}{2}\hbar\gamma\left( B_\text{L} \sigma_z^\text{L} + B_\text{R} \sigma_z^\text{R}   \right) ,
\end{align}
where $H_{\text{os}}$, $H_\text{t}$, $H_\text{Z}$ are on-site, tunneling, and Zeeman terms, respectively. Here $\tau_x$ and $\tau_z$ are Pauli matrices acting on the orbital degree of freedom in the left-right basis, e.g., $\tau_z = \ket{L}\!\!\bra{L} - \ket{R}\!\!\bra{R}$, and $\sigma_z$ is the Pauli-Z matrix acting on the electron spin.
Furthermore, $\epsilon$ denotes the on-site energy detuning between the two dots, and the (spin-conserving) tunneling amplitude is denoted by $t_0$. In the Zeeman term, $\gamma$ is the electron gyromagnetic ratio, and $B_\text{L,R}$ is the external magnetic field in the left or right dot. 

As discussed in Sec.~\ref{sec:physical-pulse-sequence}, the Hamiltonian in Eq.~\eqref{eq:HDQD}, becomes time-dependent during our Protocol, through the tunneling amplitude and the on-site energy detuning.
However, the tunneling term is spin-conserving, which means that there is no mixing between the spin-up and the spin-down sectors during the dynamics, thus the Hamiltonian has a block-diagonal structure: it consists of $2\times 2$ blocks for each spin species. Therefore the exact (time-dependent) energy eigenvalues can be obtained. We also assume that the time evolution of the idle qubits is adiabatic, i.e. they remain in their instantaneous energy eigenstate.

Let us consider an idle qubit initially located in the left dot ($t_0(0)=0$, $\epsilon(0)<0$) which is assumed to be in an even column, with Larmor frequency $\omega_e = \gamma B_\text{L}$, and the right dot is in an odd column with Larmor frequency $\omega_o = \gamma B_\text{R}$, see Fig.~\ref{fig:crossbar-setup}.
For our purposes, it is sufficient to consider the case when the detuning is large compared to the Zeeman splitting and the tunneling amplitude, $\hbar\omega_{e/o},\, t_0 \ll |\epsilon|$, and to focus on the lowest two eigenstates in energy, that is, the bonding spin-up state and the bonding spin-down state.
We expand the exact energy eigenvalues up to first order in magnetic field and second order in tunnel coupling. Thus, we obtain a $2\times 2$ effective Hamiltonian describing the idle qubit dynamics:
\begin{equation} \label{eq:effective-hamiltonian}
    H_{\text{eff}} \approx \dfrac{1}{2}\hbar\omega_e\sigma_z\left(1-\dfrac{t_0^2}{\epsilon^2}\right) + \dfrac{1}{2}\hbar\omega_o\sigma_z\dfrac{t_0^2}{\epsilon^2} ,
\end{equation}
where we omitted the time dependence of $t_0$ and $\epsilon$ for brevity.
The error mechanism described by the effective Hamiltonian Eq.~\eqref{eq:effective-hamiltonian} is that the idling qubit leaks into the neighboring dot upon barrier activation, which causes unwanted qubit rotation around $z$ (i.e., a phase gate), since the Larmor frequencies of the two dots are different.

The idle qubit located in the even column is static in a rotating frame described by 
\begin{equation}\label{eq:rotating-frame}
    W(t) = e^{\frac{i}{2} \omega_e \sigma_z  t},
\end{equation}
which transforms the Hamiltonian as
\begin{equation} \label{eq:effective-hamiltonian-rotating-frame}
    \tilde{H}_{\text{eff}} = W(t) H_{\text{eff}} W^\dagger(t)-\dfrac{\hbar}{i}\Dot{W}(t)W^\dagger(t) = -\dfrac{t_0^2}{\epsilon^2}\hbar\omega^\prime\sigma_z ,
\end{equation}
where $\omega^\prime = (\omega_e - \omega_o)/2$. 
The time-evolution operator in the rotating frame, assuming adiabatic dynamics, can be written as 
\begin{equation}
    \tilde{\mathcal{U}}_{\text{idle}} = e^{i \Delta\phi_{\text{idle}} \sigma_z /2},
\end{equation}
where the idle qubit rotation angle is
\begin{equation}\label{eq:idle-qubit-phases}
    \Delta\phi_{\text{idle}} = -2\omega^\prime \int_0^T \dfrac{t_0^2(t)}{\epsilon^2(t)} \dd t .
\end{equation}
The above calculation is applicable not only for shuttle operations: crosstalk errors during PSB readout can be characterized similarly (see Appendix~\ref{appendix:idle-qubit-errors} for more details).

Idle qubits also suffer errors due to off-resonant drive during the ESR pulses.
Such crosstalk errors occur e.g. when qubits located at the odd columns are driven with ESR pulse outlined in Eq.~\eqref{eq:ESR-drive-pulse}, using $\omega_{\text{ac}} = \omega_o$, meanwhile, qubits located in the even columns are idling (see Steps~\ref{step:global-H},~\ref{step:ancilla-H} and \ref{step:ancilla-H-2} in our Protocol). In this case, idle qubits are subjected to unwanted partial Rabi oscillations. 
We estimate the average error they suffer using the infidelity formula
\begin{equation}\label{eq:avg-fidelity}
    1-F =\dfrac{2}{3} - \dfrac{1}{6} \Big|\Tr\left(\mathcal{U}_0^\dagger \mathcal{U}\right)\Big|^2,
\end{equation}
where $F$ denotes the average fidelity (averaged over all possible idle qubit initial state on the Bloch sphere) \cite{Pedersen_2007}, $\mathcal{U}_0$ is the $2\times 2$ identity matrix, and $\mathcal{U}$ is the time-evolution operator in the idle qubit frame [defined in Eq.~\eqref{eq:rotating-frame}]. Starting from the lab frame Hamiltonian
\begin{equation}
    H_e = \dfrac{1}{2}\hbar\omega_e \sigma_z + \Omega\sin\left(\omega_o t + \varphi\right) \sigma_x,
\end{equation}
we calculate the time-evolution operator employing the rotating wave approximation in the drive frame [defined by substituting $\omega_e$ with $\omega_o$ in Eq.~\eqref{eq:rotating-frame}], and then transforming that result to the qubit frame.
For large detuning $\omega^\prime\gg\Omega$, Eq.~\eqref{eq:avg-fidelity} yields
\begin{equation} \label{eq:idle-error-ESR} 
    1-F\lesssim \alpha \dfrac{\Omega^2}{\left(\omega^\prime\right)^2},
\end{equation}
where $\alpha$ is a constant, e.g. for an off-resonant $X$ gate ($\varphi = \pi /2$, and $t=\pi / \Omega$) it is $\alpha = (64 + \pi^2) / 384$.
For an off-resonant Hadamard gate composed of two consecutive single-qubit rotations [see Eq.~\eqref{eq:hadamard}], the idle qubit error is roughly doubled, $2\alpha \Omega^2 / \left(\omega^\prime\right)^2$.

In Appendix~\ref{appendix:idle-qubit-errors}, Table~\ref{tab:idle-qubit-rot} provides an exhaustive list of unwanted idle-qubit rotations together with the estimated errors based on the arguments we discussed here, using a realistic set of parameters (Table~\ref{tab:simulation-parameters}). Since none of the estimated idle qubit errors exceeds the per-step threshold error probability of the surface code, $p_{\text{th}} = 0.57\%$ \cite{Fowler_2012}, we expect that the suppression of logical errors remains intact. 

We also note that due to the control constraints of the crossbar architecture, the spatial distribution of idle qubit errors has a periodic structure. This implies inhomogeneous (but spatially periodic) coherent noise on the surface code logical state which in principle could have an enhanced effect compared to random Pauli errors. However recent works showed that for the surface code, coherent single-qubit rotations do not change the error correcting threshold significantly \cite{Bravyi_2018,Marton2023coherenterrors,Pataki2024coherenterrors}, supporting the validity of our simple error estimates.

A refined version of the estimated error scaling described in Sec.~\ref{sec:error-scaling} can be obtained by using the simulated shuttling-based Z gate time duration, $\tau_z^{\prime} \approx 10\unit{ns}$ (with simulation parameters listed in Table~\ref{tab:simulation-parameters}) and including the idle qubit errors in the average error probability, Eq.~\eqref{eq:P-tot-param}. Based on Table~\ref{tab:idle-qubit-rot}, we estimate the average idle qubit error for the Z cycle as $p_{\text{idle}} \approx 6.4\cdot 10^{-4}$ and the average number of imperfect idle qubit operations for a full cycle to be $8.5$ per qubit. The latter is obtained as the sum of the number of idle qubit errors in each step, divided by the total number of qubits in the unit cell (see Fig.~\ref{fig:unit-cell}). 

\begin{figure}
    \centering
    \includegraphics[width=0.48\textwidth]{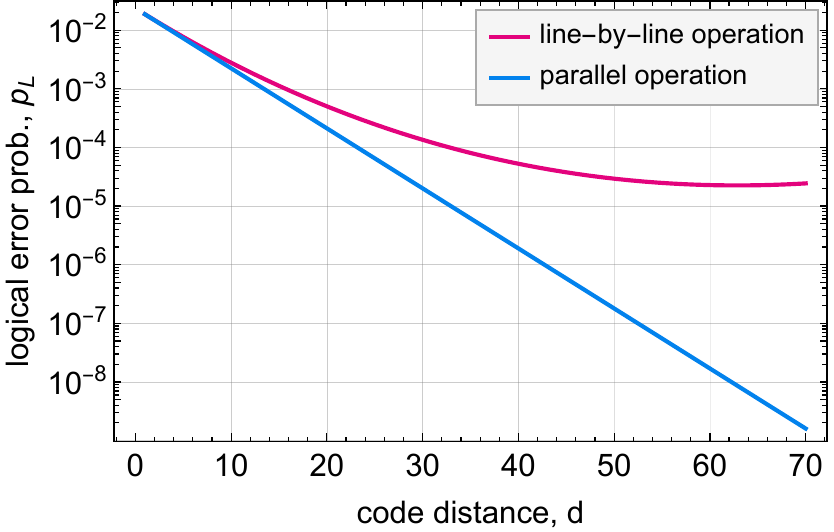}
    \caption{Refined estimate of the logical error probabilities versus the code distance, based on Eq.~\eqref{eq:logical-error-rate}, with a shorter Z-rotation time ($\tau_z^{\prime} \approx 10\unit{ns}$), and accounting for idle qubit errors as well. Plotted for line-by-line and parallel operations, using coherence time $T_2^\ast = 10\unit{ms}$ and measurement time $\tau_m = 1000\unit{ns}$. Parallel operation provides better logical performance (with exponential error suppression factor $\Lambda\approx 1.6$), meanwhile, logical error suppression also remains feasible for line-by-line operation in this case, for moderate code sizes, up to $d_{\text{opt}} = 63$, and with the lowest attainable logical error probability being $p_L \approx 2.3\cdot 10^{-5}$.
    }
    \label{fig:refined-logical-error-rate}
\end{figure}

Assuming that in the X cycle, the average idle qubit error is approximately the same, this gives rise to an additional term, $17 p_{\text{idle}}$ in Eq.~\eqref{eq:P-tot-param}, which is a dominant contribution to the average error probability. Taking into account this additional error term and substituting $\tau_z$ with $\tau_z^{\prime}$ in Eqs.~\eqref{eq:tau-tot-line-by-line}-\eqref{eq:tau-tot-parallel}, we obtain a refined estimate for the logical error probabilities, shown in Fig~\ref{fig:refined-logical-error-rate}. These results show that in the presence of idle qubit errors, parallel operation has decreased exponential error suppression, $\Lambda\approx 1.6$. Since the idle qubit errors are small, logical error suppression remains feasible for the line-by-line implementation as well, up to moderate code sizes, $d_{\text{opt}} = 63$, with the lowest attainable logical error probability being $p_L \approx 2.3\cdot 10^{-5}$.

\section{Discussion}
\label{sec:discussion}

\textit{Error scaling for an inhomogeneous (non-uniform) crossbar system.} In this work, we considered a uniform (perfect) crossbar system only. However, in real experiments, quantum hardware always has some imperfections. A natural extension would be to incorporate such inhomogeneities in our logical error estimate.

\textit{Decoding and error correction.}
In a quantum error correction experiment, the simplest option is to process the measured data after the syndrome measurement circuits are executed on the quantum processor \cite{GoogleQuantum_2024}. 
Having access to a fast and accurate decoder \cite{Pattison2021ImprovedQE,bausch2023learningdecodesurfacecode,shutty2024efficientnearoptimaldecodingsurface,Namitha2024UnionFind,jones2024improvedaccuracydecodingsurface} and feed-forward operations, the correction can be performed in real-time, allowing for more sophisticated fault-tolerant protocols as well.
In the crossbar architecture, Z errors (e.g. crosstalk errors) could be corrected using shuttling-based $z$ rotations, acquiring significantly shorter duration than semi-global rotations.

\section{Conclusions}
\label{sec:conclusions}

In this work, we have made important steps towards the physical realization of quantum circuits in crossbar spin qubit architectures. 
In particular, we have identified an abstract pulse sequence, as well as a corresponding physical pulse sequence, which realizes surface-code quantum error correction in this platform.
As an important ingredient, we have developed a verification algorithm that confirms that the gate-voltage changes in the abstract pulse sequence induce the desired real-space routing of the electrons in the quantum dot array.
The identification of the physical pulse sequence enabled us to refine the estimation procedure of the logical error probability.
In fact, we show a concrete, realistic parameter set that enables below-threshold surface code operation.
The concepts and methods introduced here facilitate the realisation of quantum computing and quantum error correction with crossbar spin qubit architectures.

\section*{Acknowledgments}

We thank P.~Boross, Z.~Guba, B.~Kolok and J.~Helsen for helpful discussions. 
This research was supported by the Ministry of Culture and Innovation and the National Research, Development and Innovation Office within the Quantum Information National Laboratory of Hungary (Grant No. 2022-2.1.1-NL-2022-00004). 
This research has been supported by the Horizon Europe research and innovation programme of the European Union through the projects IGNITE and QLSI2, as well as by the National Research, Development and Innovation Office via the OTKA Grant No. 132146.
This project has received funding from the HUN-REN Hungarian Research Network.

\appendix

\section{Calculation details for the error scaling with line-by-line implementation} \label{appendix:line-by-line}

This Appendix contains four Tables supporting various parts of the main text. 
In Table~\ref{tab:list-of-corrections}, we collect the six corrections we propose compared to Ref.~\cite{Helsen_2018}.
These corrections are incorporated in our analysis in Sec.~\ref{sec:error-scaling}.

Calculation details for the error scaling with line-by-line implementation are summarized in Tables~\ref{tab:z-cycle-time-step-counts}-\ref{tab:gate-count}. Namely, Table~\ref{tab:z-cycle-time-step-counts} lists the composite time-step count of each Step of our Protocol in terms of the types of operations for the line-by-line implementation of the Z parity checks (described in Sec.~\ref{sec:line-by-line}). Table~\ref{tab:total-time-step-count} summarizes the total number of composite time steps for all gate types for a full surface code stabilizer measurement cycle. Similar considerations hold for parallel operation as well, substituting $d=1$ throughout Table~\ref{tab:total-time-step-count}. We use these results in the main text to estimate the time duration of the full stabilizer measurement cycle, see Eqs.~\eqref{eq:tau-tot-line-by-line} and \eqref{eq:tau-tot-parallel}.

We also list the number of operations per qubit type (data/ancilla qubit) for both cycles in Table~\ref{tab:gate-count}, and we use this data for the estimate of the average error probability per stabilizer measurement cycle, in Eq.~\eqref{eq:P-tot-param}.

\begingroup
\setlength{\tabcolsep}{5pt}
\renewcommand{\arraystretch}{1.25}
\begin{table*}
    \centering
    \begin{tabular}{c p{0.65\linewidth} c c c }
    \hline\hline
        No. & \centering Result & Ref.~\cite{Helsen_2018} &  Our version \\ \hline
        1 & Decomposition of the CNOT gate in terms of $\sqrt{\text{SWAP}}$ gates & Fig.~4 & Fig.~\ref{fig:cnot-with-sqrt-swaps}  \\
        2 & Surface code Z-stabilizer measurement circuit decomposed with two-qubit $\sqrt{\text{SWAP}}$ gates and single-qubit gates & Fig.~11 & Fig.~\ref{fig:z-cycle-with-sqrt-swaps}    \\
        3 & Time step count per step in terms of gate types for the line-by-line implementation of the surface code stabilizer measurement cycle & Table~4 & Table~\ref{tab:z-cycle-time-step-counts} \\
        4 & Number of operations per qubit type (data/ancilla qubit) for the X and Z stabilizer measurement cycles & Table~5 & Table~\ref{tab:gate-count} \\
        5 & Estimated time of a full stabilizer measurement cycle using line-by-line operation & Eq.~(18) & Eq.~\eqref{eq:tau-tot-line-by-line} \\
        6 & Average physical error probability per quantum stabilizer measurement cycle  & Eq.~(20) & Eq.~\eqref{eq:P-tot-param} \\ \hline\hline
    \end{tabular}
    \caption{List of corrections (1-2) and modifications (3-6) we propose compared to Helsen \textit{et al.} \cite{Helsen_2018}.}
    \label{tab:list-of-corrections}
\end{table*}
\endgroup

\begingroup
\setlength{\tabcolsep}{5pt}
\begin{table*}
    \centering
    \begin{tabular}{cccccccccccccccc}
    \hline\hline
        Steps & \ref{step:init} & \ref{step:global-H} & \ref{step:right-triangle} & \ref{step:ancilla-H} & \ref{step:sqrt-of-swaps} & \ref{step:back-to-idle} & \ref{step:left-triangle} & \ref{step:sqrt-of-swaps-2} & \ref{step:ancilla-H-2} & \ref{step:back-to-idle-2} & \ref{step:global-H-2} & \ref{step:measurement-conf} & \ref{step:readout} & \ref{step:measurement-to-idle-conf} & Z cycle total \\ \hline
        $\sqrt{\text{SWAP}}$ gates &  &  &   &  & $4d$ &  &  & $4d$ &  &  &  &  &  &  & $8d$  \\
        $Z$-rotations             &  &  &  &  & $4d$ &  &  & $4d$ &  &  &  &  &  &  & $8d$  \\
        Shuttling                 &  &   & $d$ &  &  & $d$ & $d$ &  & & $d$ &  & $d$ &  & $d$ & $5d$ \\
        Global rotations          &  & $1$  &  & $1$&  & &  &  & $1$ &  & $1$ &  &  &  & $4$ \\
        Measurements               &  &  &  &  &  & &  &  &  &  &  &  & $d$ &  & $d$ \\\hline\hline
    \end{tabular}
    \caption{Composite time step count per Step in terms of gate types for the line-by-line implementation of the surface code Z stabilizer measurement cycle. $Z$-rotations refers to shuttling-based single-qubit rotations around the z-axis. Table cells that are left empty signify zero entries. 
    For the X stabilizer measurement cycle, we need one more global rotation and one less shuttling-based $Z$-rotation. A similar table could be obtained for parallel operation, substituting $d=1$ into the $d$-dependent entries of this Table.
    }
    \label{tab:z-cycle-time-step-counts}
\end{table*}
\endgroup

\begin{table}
    \centering
    \begin{tabular}{ccc}
    \hline\hline
        Operation                 & Time steps per full cycle &\,  Notation \\ \hline
        $\sqrt{\text{SWAP}}$ gate & $16d$ & $\tau_{sw}$ \\
        $Z$-rotation              & $15d$ & $\tau_{z}$ \\
        Shuttling                 & $10d$ & $\tau_{sh}$ \\
        Global rotation           & $9$ & $\tau_{gl}$ \\
        Measurement               & $2d$ & $\tau_{m}$ \\ \hline\hline
    \end{tabular}
    \caption{Total number of composite time steps per operation for the full surface code stabilizer measurement cycle (subsequent X and Z stabilizer measurements) and the corresponding notations, assuming line-by-line operation. A similar table could be obtained for parallel operation, substituting $d=1$ into the $d$-dependent entries of this Table.
    }
    \label{tab:total-time-step-count}
\end{table}

\begingroup
\setlength{\tabcolsep}{4pt}
\begin{table*}
    \centering
    \begin{tabular}{cccccccccccccc}
    \hline\hline
          & \multicolumn{3}{c}{Data qubits} && \multicolumn{3}{c}{Z ancilla qubits} && \multicolumn{3}{c}{X ancilla qubits} && \multirow{2}{*}{Average data/ancilla} \\ \cline{2-4}\cline{6-8}\cline{10-12}
          & Z-cycle   & X-cycle   & Total   && Z-cycle    & X-cycle    & Total  && Z-cycle    & X-cycle    & Total  &&                          \\ \hline
       $\sqrt{\text{SWAP}}$ gate  & 2 & 2 & 4 && 8 & 0 & 8 && 0 & 8 & 8 && 6 \\
       $Z$-rotation               & 1 & 0 & 1 && 4 & 0 & 4 && 0 & 7 & 7 && 3.25 \\
       Shuttling                  & 0 & 0 & 0 && 6 & 0 & 6 && 0 & 6 & 6 && 3 \\
       Global rotation            & 1 & 3 & 4 && 2 & 2 & 4 && 2 & 3 & 5 && 4.25 \\
       Measurement                & 0 & 0 & 0 && 1 & 0 & 1 && 0 & 1 & 1 && 0.5 \\ \hline\hline
    \end{tabular}
    \caption{Number of operations per qubit type (data/ancilla qubit) for both cycles. Average values in the last column are obtained as the weighted average operation number for data qubits and ancilla qubits, using weight 2 for the data qubits. Qubits located at the boundary of the code might have lower gate counts.}
    \label{tab:gate-count}
\end{table*}
\endgroup

\section{Estimation of idle qubit errors} \label{appendix:idle-qubit-errors}

In this appendix, we derive and estimate the unwanted idle qubit rotations described in Sec.~\ref{sec:idle-errors}. In the rotating frame, using the effective Hamiltonian, Eq.~\eqref{eq:effective-hamiltonian-rotating-frame} in the main text, we obtain the idle qubit rotation angles as per Eq.~\eqref{eq:idle-qubit-phases},
where the on-site energy detuning between the left and the right dot is $\epsilon(t) = \varepsilon_{L}(t) - \varepsilon_{R}(t)$.
When the time dependence of the on-site energies and the tunneling amplitude are described by Eqs.~\eqref{eq:plunger-reset} and \eqref{eq:barrier-activation}, the integral can be carried out exactly. Thus, crosstalk during shuttling results in idle qubit rotation angle
\begin{widetext}
\begin{equation} \label{eq:idle-error-shuttling} 
    \Delta\phi_{\text{idle}}^{(\text{shuttle})} = \frac{12 \left(A^2-4 A \left(\sqrt{\epsilon_0 (\epsilon_0-A)}+2 \epsilon_0\right)+8 \epsilon_0 \left(\sqrt{\epsilon_0 (\epsilon_0-A)}+\epsilon_0\right)\right)}{A^4}  t_{0,\text{max}}^2 \omega^\prime T ,     
\end{equation}
where $\epsilon_0 = u_{L,i} - u_{R,i}$ is the initial detuning between the dots which is assumed to have a large negative value ($\epsilon_0 \ll 0$), and
\begin{equation}
    A = u_{L,j}+u_{R,j}-u_{L,i}-u_{R,i} 
\end{equation}
is the detuning amplitude, which has a smaller magnitude, $|A|<|\epsilon_0|$, so that the overall detuning remains negative throughout the protocol. 

Similarly, for shuttling-based $z$ rotations, using the gate voltage and tunnel barrier pulses of Eq.~\eqref{eq:shuttling-based-z-rot}, we obtain
\begin{equation} \label{eq:idle-error-z-rot}
    \Delta\phi_{\text{idle}}^{(\text{z-rot.})} = \frac{ \left(3 A^\prime \epsilon_0^\prime-2 \epsilon_0^\prime \left(\sqrt{\epsilon_0^\prime (\epsilon_0^\prime-A^\prime)}+\epsilon_0^\prime\right)+2 A^\prime \sqrt{\epsilon_0^\prime (\epsilon_0^\prime-A^\prime)}\right)}{(A^\prime)^2 \sqrt{-\epsilon_0^\prime} (A^\prime-\epsilon_0^\prime)^{3/2}} t_{0,\text{max}}^2 \omega^\prime T_z , 
\end{equation}
\end{widetext}
where we use the notation $\epsilon_0^\prime = u_{L,\text{t}} - u_{R,\text{t}} <0$ and 
\begin{equation}
    A^\prime = u_{L,\text{n}}+u_{R,\text{n}}-u_{L,\text{t}}-u_{R,\text{t}} .
\end{equation}
Note that in the special case, when the two on-site energies $\varepsilon_L$ and $\varepsilon_R$ are tuned such that the detuning $\epsilon(t)$ is constant in time ($A=0$), Eq.~\eqref{eq:idle-error-shuttling} yield a simplified form,
\begin{equation} \label{eq:idle-error-simple}
    \Delta\phi_{\text{idle}} = \dfrac{3 t_{0,\text{max}}^2}{4 \epsilon_0^2}  \omega^\prime T .
\end{equation} 
Eq.~\eqref{eq:idle-error-z-rot} also yields a similar form when $A^\prime =0$.

Idling data qubits during the PSB readout (depicted in Fig.~\ref{fig:parallel-measurement}) can also suffer such unwanted rotations. We calculate these rotation angles similarly, using the effective Hamiltonian, Eq.~\eqref{eq:effective-hamiltonian-rotating-frame}. During the readout, the vertical barrier between the ancilla qubits is lowered and the on-site energy of the reference qubit (blue qubits in Fig.~\ref{fig:parallel-measurement}) is lowered such that the attained detuning $\epsilon_{\text{max}}$ is large enough to overcome the on-site Coulomb repulsion.
However, for the readout it is sufficient to activate the barrier only when the on-site energy detuning is approaching the value of the on-site Coulomb repulsion $U$, reducing the unwanted rotation angles significantly. Therefore, preceded by an abrupt plunger reset, we simulate the spin-to-charge conversion as per Eqs.~\eqref{eq:plunger-reset} and~\eqref{eq:barrier-activation}, with reference qubit on-site energies $u_i = -2100\unit{\mu eV}$, $u_j = -2500\unit{\mu eV}$, and time duration $T=25\unit{ns}$. 
Meanwhile, the on-site energy of the measured qubit (red qubits in Fig.~\ref{fig:parallel-measurement}), is constant at $u_1 = -800\unit{\mu eV}$.

From the rotation angles, we estimate idle qubit errors as the maximal infidelity with the initial state.
In the case of any shuttling-based operation (when there is barrier activation), idle qubits can suffer $z$ rotations as crosstalk errors. Rotations around the $z$ axis are the most harmful for states in an equal superposition of computational basis states, such as 
\begin{equation} 
    \ket{\tilde{\psi}_0} = \dfrac{1}{\sqrt{2}} \left( \ket{\tilde{0}} + \ket{\tilde{1}} \right),
\end{equation}
 
Thus we obtain the idle error as
\begin{equation} \label{eq:idle-infidelity}
   1- \Big|\mel**{\tilde{\psi}_0}{\tilde{\mathcal{U}}_{\text{idle}}}{\tilde{\psi}_0} \Big|^2 = \sin^2\left(\Delta\phi_{\text{idle}} /2\right).
\end{equation}
We note that using the average infidelity formula, Eq.~\eqref{eq:avg-fidelity}, we get a very similar result, 
\begin{equation} \label{eq:idle-infidelity}
   1- F = \dfrac{2}{3}\sin^2\left(\Delta\phi_{\text{idle}} /2\right),
\end{equation}
that would provide a slightly more optimistic estimate.

The obtained idle qubit errors are summarized in Table~\ref{tab:idle-qubit-rot}. For the numerical values listed in the last column we used the model parameters listed in Table~\ref{tab:simulation-parameters}. are all below the surface code per-step error threshold $p_{\text{th}} = 0.57\%$ \cite{Fowler_2012}, therefore we expect that these crosstalk errors are tolerated by the code, at least for parallel operation (see Fig.~\ref{fig:refined-logical-error-rate}). We also note that the idle qubit rotations could be suppressed by using a lower tunnel amplitude $t_{0,\text{max}}$ [see Eqs.~\eqref{eq:idle-error-shuttling},\eqref{eq:idle-error-simple},\eqref{eq:idle-error-z-rot}].

\begin{table*}
    \centering
    \begin{tabular}{ccccc}
    \hline\hline
        Operation                 &  Corresponding steps & Idle qubits & Formula & Estimated error\\ \hline

        \multirow{2}{*}{idle configuration $\longleftrightarrow$ leftward/rightward triangle}   &\, Steps~\ref{step:right-triangle},~\ref{step:back-to-idle},~\ref{step:left-triangle},~\ref{step:back-to-idle-2},  \ref{step:measurement-conf},~\ref{step:measurement-to-idle-conf} &\, D1, D2, D3, D4 & Eq.~\eqref{eq:idle-error-shuttling} &  $1.8\cdot 10^{-4}$\\
           & Steps~\ref{step:right-triangle},~\ref{step:back-to-idle-2}  & B1, B2 & Eq.~\eqref{eq:idle-error-simple} & $2.7\cdot 10^{-5}$  \\[0.5em]

        \multirow{2}{*}{shuttling-based $Z$ gates}   & \multirow{2}{*}{Steps~\ref{step:sqrt-of-swaps},~\ref{step:sqrt-of-swaps-2}} & D1, D2, D3, D4 & Eq.~\eqref{eq:idle-error-simple} & $2.3\cdot 10^{-3}$\\
           &  & A1, A2 & Eq.~\eqref{eq:idle-error-z-rot} & $2.5\cdot 10^{-3}$\\[0.5em]

        \multirow{2}{*}{shuttling-based $S^\dag$ gates}   & \multirow{2}{*}{Steps~\ref{step:sqrt-of-swaps},~\ref{step:sqrt-of-swaps-2}} & D1, D4 & Eq.~\eqref{eq:idle-error-simple} & $7.3\cdot 10^{-5}$ \\
           &  & D2, D3 & Eq.~\eqref{eq:idle-error-z-rot} & $7.1\cdot 10^{-4}$\\[0.5em]

        \multirow{2}{*}{PSB readout}   & \multirow{2}{*}{Step~\ref{step:readout}} & D1, D4 & Eq.~\eqref{eq:idle-error-simple} & $2.0\cdot 10^{-3}$\\
           &  & D2, D3 & Eq.~\eqref{eq:idle-error-shuttling} & $5.7\cdot 10^{-6}$\\[0.5em]

        \multirow{3}{*}{Hadamard gate}  & Step~\ref{step:global-H} & D1, D2, D3, D4 & Eq.~\eqref{eq:idle-error-ESR} & $2.2\cdot 10^{-4}$ \\
           & \multirow{2}{*}{Step~\ref{step:ancilla-H},~\ref{step:ancilla-H-2}} & D1, D2, D3, D4 & \multirow{2}{*}{Eq.~\eqref{eq:idle-error-ESR}} & \multirow{2}{*}{$2.2\cdot 10^{-4}$} \\
           &  & A1, A2 &  & \\[0.5em]
        
        \hline\hline
    \end{tabular} 
    \caption{List of idle qubit errors during the surface code Z-cycle. Qubit notations are as per Fig.~\ref{fig:unit-cell}. We used the parameters listed in Table~\ref{tab:simulation-parameters} for the error estimation. From the rotation angles expressed by the equations in column "Formula", we estimate the errors using the infidelity, Eq.~\eqref{eq:idle-infidelity}. Our estimates provide approximate upper bounds for the actual idle qubit errors in each step. Note that ancilla qubits B1 and B2 are not affected between Steps~\ref{step:ancilla-H} and \ref{step:ancilla-H-2} (and after Step~\ref{step:global-H-2}), since after two (or four) subsequent Hadamard gates they are ideally in a computational basis state. The average (weighted with the number of steps and the number of qubits) estimated idle qubit error is  $p_{\text{idle}} \approx 6.4\cdot 10^{-4}$ and the average number of imperfect idle qubit operations for a full cycle is 17 per qubit.}
    \label{tab:idle-qubit-rot}
\end{table*}

\begin{table*}
    \centering
    \begin{tabular}{ccc}
    \hline\hline
        Parameter                       &\, Notation &\,  Value \\ \hline
        Maximal tunneling amplitude            &\, $t_{0,\text{max}}$ &\, $25\unit{\mu eV}$ \\
        Larmor frequency difference     &\, $\omega^\prime = (\omega_e - \omega_o)/2$ &\, $1\unit{\mu eV} / \hbar$ \\
        On-site energy level 1          & $u_1$ & $-800 \unit{\mu eV}$ \\
        On-site energy level 2          & $u_2$ & $-533.33 \unit{\mu eV}$ \\
        On-site energy level 3          & $u_3$ & $-266.66 \unit{\mu eV}$ \\
        On-site energy level 4          & $u_4$ & $0\unit{\mu eV}$ \\
        On-site repulsion               & $U$ & $1500\unit{\mu eV}$ \\
        On-site energy detuning for PSB readout  & $\epsilon_{\text{max}}$ & $1700\unit{\mu eV}$ \\
        Shuttling time                  & $\tau_{sh}$ & $9.35\unit{ns}$ \\
        Time duration of shuttling-based $Z$ gate       & - & $9.62\unit{ns}$ \\
        Time duration of shuttling-based $S^\dag$ gate  & - & $4.82\unit{ns}$ \\ 
        PSB spin-to-charge conversion time               & -   & $25\unit{ns}$  \\
        ESR drive amplitude  & $B_{\text{ac}}$ & $0.35\unit{mT}$ \\
        Qubit Rabi frequency  & $\Omega$ & $0.02\unit{\mu eV} / \hbar$ \\ \hline\hline
    \end{tabular} 
    \caption{List of parameters used for the simulation of physical pulses based on Sec.~\ref{sec:physical-pulse-sequence} and \ref{sec:idle-errors}. Parameters are chosen to ensure that the target qubit transfer during shuttling is adiabatic. Time durations of shuttling, the $Z$ and the $S^\dag$ gates are obtained by numerical optimization of the corresponding target qubit fidelity while other parameters were fixed.}
    \label{tab:simulation-parameters}
\end{table*}

\begin{figure}
    \centering
    \includegraphics[width=0.48\textwidth]{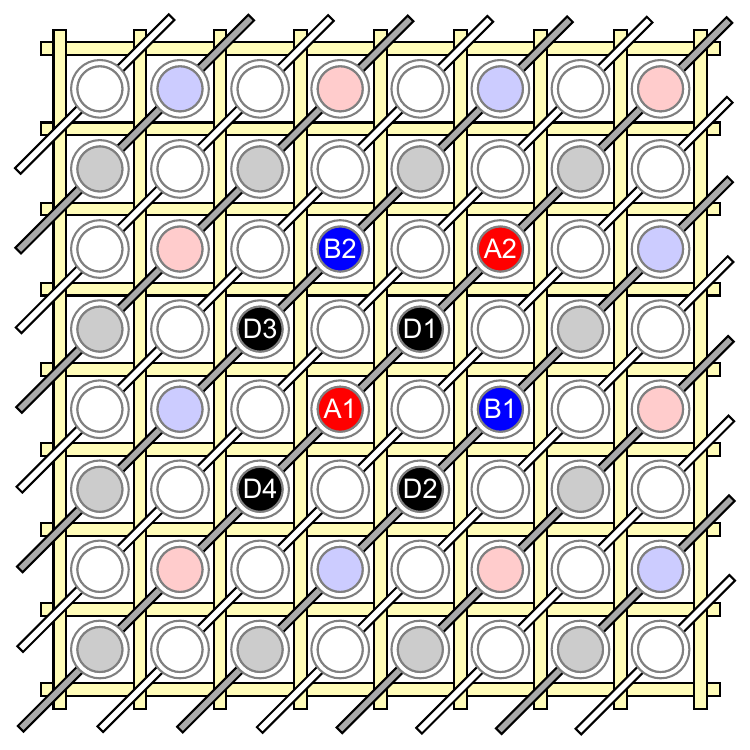}
    \caption{Data qubits and ancilla qubits in the idle configuration. The $4\times 4$ unit cell of the grid is highlighted and the corresponding 8 qubits in this cell are numbered.}
    \label{fig:unit-cell}
\end{figure}

\section{Short guide to Supplementary Material files} \label{appendix:suppmat-files}

\begin{figure}
    \centering
    \includegraphics[width=0.42\textwidth]{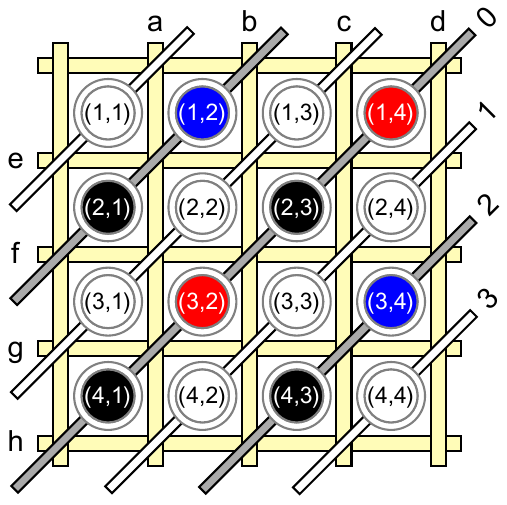}
    \caption{The $4\times 4$ unit cell of the grid in the idle configuration with qubit coordinates and new notations for the barriers used in auxiliary files shiftsZ.txt and shiftsX.txt.}
    \label{fig:notations-in-unit-cell}
\end{figure}

Here, we give an overview of the Supplementary Material files \cite{suppmat_Zenodo}. Apart from the present manuscript, our work includes the following auxiliary files:
\begin{enumerate}
    \item ZcyclePulseSequence.xlsx: Excel sheet containing the abstract gate voltage and ESR pulse sequence for the surface code Z-cycle, as described in Sec.~\ref{sec:abstract-pulse-sequence}, outlined in Table~\ref{tab:abstract-pulse-sequence}.
    \item XcyclePulseSequence.xlsx: Excel sheet containing the abstract gate voltage and ESR pulse sequence for the surface code X-cycle, as described in Sec.~\ref{sec:abstract-pulse-sequence}.
    \item configurationsZ.txt: Text file containing the configurations (electron configurations and plunger configurations, as defined in Sec.~\ref{sec:shuttling-protocol}, see item~\ref{configuration}) whose sequence implements the Z-cycle shuttling protocol, ($\mathtt{C}_0$, $\mathtt{C}_1$,~\dots, $\mathtt{C}_6$) = ($(\mathtt{E}_0,\mathtt{P}_0)$, $(\mathtt{E}_1,\mathtt{P}_1)$,~\dots, $(\mathtt{E}_6,\mathtt{P}_6)$), as rows. 
    Qubit coordinates are as per Fig.~\ref{fig:notations-in-unit-cell}. This file is an input for the shuttling verification algorithm.
    \item shiftsZ.txt: Text file containing the shifts (describing the movement of electrons, as defined in Sec.~\ref{sec:shuttling-protocol}, see item~\ref{shift}) of the Z-cycle shuttling protocol, ($\mathtt{S}_{0\to 1}$, $\mathtt{S}_{1\to 2}$,~\dots, $\mathtt{S}_{5\to 6}$) = ($(\beta_{0\to 1},\mathtt{P}_{0\to 1})$, $(\beta_{1\to 2},\mathtt{P}_{1\to 2})$,~\dots, $(\beta_{5\to 6},\mathtt{P}_{5\to 6})$), as rows. Here, we use a slightly modified notation for convenience, depicted in Fig.~\ref{fig:notations-in-unit-cell}. This file is an input for the shuttling verification algorithm.
    \item configurationsX.txt: Text file containing the electron configurations (defined in Sec.~\ref{sec:shuttling-protocol}) whose sequence implements the X-cycle shuttling protocol.  Qubit coordinates are as per Fig.~\ref{fig:notations-in-unit-cell}. This file is an input for the shuttling verification algorithm.
    \item shiftsX.txt: Text file containing the shifts (describing the movement of electrons) of the X-cycle shuttling protocol, as rows. Here, we use a slightly modified notation for convenience, depicted in Fig.~\ref{fig:notations-in-unit-cell}. This file is an input for the shuttling verification algorithm.
    \item ZcycleMovie.pdf: Document containing a visual representation of the abstract pulse sequence of the surface code Z-cycle (ZcyclePulseSequence.xlsx).
    \item XcycleMovie.pdf: Document containing a visual representation of the abstract pulse sequence of the surface code X-cycle (XcyclePulseSequence.xlsx).
\end{enumerate}

Additionally, our implementation of the shuttling verification algorithm (as described in Sec.~\ref{sec:shuttling-protocol}) is available at the Zenodo repository~\cite{Pataki_Verification_of_shuttling}.

%\clearpage
\bibliography{QEC-spin-qubits}

\end{document}